\documentclass[12pt,a4paper]{article}
\pdfoutput=1

\usepackage{jheppub}
\usepackage{bm}
\usepackage{amsmath}
\usepackage{amssymb}
\usepackage{amsthm}
\usepackage{ccaption}
\usepackage{subcaption}
\usepackage{graphicx}
\usepackage{sidecap}
\usepackage[super]{nth}
\usepackage{microtype}
\usepackage{cleveref}
\usepackage{overpic}

\graphicspath{{Figures/}}

  \captionnamefont{\bfseries}
  \captiontitlefont{\small\sffamily}
  \captiondelim{: }
  \hangcaption

\def\rp{r_+}
\def\rm{r_-}
\def\length{\ell}
\def\RR{\mathbb{R}}
\def\ZZ{\mathbb{Z}}
\def\PP{\mathbb{P}}
\def\CC{\mathbb{C}}
\DeclareMathOperator{\Tr}{Tr}
\DeclareMathOperator{\csch}{csch}

\title{Entanglement entropy in three dimensional gravity}

\author{Henry Maxfield}

\affiliation{ Centre for Particle Theory \& Department of Mathematical Sciences,\\
Durham University, South Road, Durham DH1 3LE, UK.}
\emailAdd{h.d.maxfield@durham.ac.uk}

\abstract{
The Ryu-Takayanagi (RT) and covariant Hubeny-Rangamani-Takayanagi (HRT) proposals relate entanglement entropy in CFTs with holographic duals to the areas of minimal or extremal surfaces in the bulk geometry. We show how, in three dimensional pure gravity, the relevant regulated geodesic lengths can be obtained by writing a spacetime as a quotient of $AdS_3$, with the problem reduced to a simple purely algebraic calculation. We explain how this works in both Lorentzian and Euclidean formalisms, before illustrating its use to obtain novel results in a number of examples, including rotating BTZ, the $\RR\PP^2$ geon, and several wormhole geometries. This includes spatial and temporal dependence of single-interval entanglement entropy, despite these symmetries being broken only behind an event horizon. We also discuss considerations allowing HRT to be derived from analytic continuation of Euclidean computations in certain contexts, and a related class of complexified extremal surfaces.}

\arxivnumber{1412.0687}

\begin{document}

\begin{flushright} \small{DCPT-14/67} \end{flushright}

\maketitle

\flushbottom
\renewcommand{\thefootnote}{\arabic{footnote}}

\section{Introduction}

A central feature of quantum systems, distinguishing them from classical ones, is entanglement. In holographic theories, there is an intimate relationship between this quantum mechanical property and geometry. One concrete realisation of this relationship is embodied in the Ryu-Takayanagi (RT) prescription \cite{Ryu:2006ef,Ryu:2006bv} and its covariant Hubeny-Rangamani-Takayanagi (HRT) generalisation \cite{Hubeny:2007xt}, which equates the entanglement entropy of a spatial region in the field theory to the area of an extremal surface in the bulk geometry. The entanglement entropy is a natural and useful quantity in field theory, but is notoriously difficult to compute in most circumstances. The gravitational calculation is in many cases much more tractable, so offers a practical way to study the properties and dynamics of entanglement.

Three dimensional gravity has no local degrees of freedom, which makes it a particularly useful setting in which to obtain analytic results. One way of understanding this is that the Ricci tensor specifies the Riemann tensor completely, a consequence of which is that solutions of pure gravity with negative cosmological constant are necessarily isometric to pure $AdS_3$ locally. Despite this, there is still a rich set of solutions that differ from $AdS$ globally, obtained by taking a quotient of a subset of $AdS_3$ by some group of isometries. The simplest such solution is the BTZ black hole \cite{Banados:1992wn,Banados:1992gq}, later generalised to `wormholes' with any number of asymptotic boundaries, joined through regions with possibly nontrivial topology \cite{Brill:1995jv,Aminneborg:1997pz,Brill:1998pr}, and possibly rotating \cite{Aminneborg:1998si}. Further generalisations include non-orientable manifolds, see \cite{Yin:2007at} for example. In many cases, there is a known candidate to construct the CFT state dual to these geometries \cite{Skenderis:2009ju,Krasnov:2000zq}, generalising the interpretation of an eternal black hole as a thermofield double state \cite{Maldacena:2001kr}.

A priori, the calculation of entanglement entropies, even with holography and with the simplifications of three dimensions, appears challenging in these geometries. It requires finding geodesics, imposing an appropriate infrared cutoff and then integrating their lengths, possibly in a space with complicated topology and no symmetry, and requiring several coordinate patches. The main result of this paper is to show that this apparent difficulty can be circumvented, and in fact the geodesic lengths can be found purely from algebra. With a description of a spacetime as a quotient in hand, the lengths are computed simply from the trace of a product of $2\times2$ matrices, representing points on the boundary and elements of the quotient group.

From a computational point of view, the advantages of this approach are obvious: many lengths can be calculated very quickly and algorithmically, allowing for very detailed investigations of entanglement entropy in these geometries. We demonstrate this utility in \cref{sec:examples}, by computations in several examples. We begin for illustration by reproducing known results in rotating BTZ, a single line computation once the black hole has been written as a quotient. We then move on to the $\RR\PP^2$ geon to demonstrate the generalisation to nonorientable spacetimes, and describe the structure of entanglement in the associated pure CFT state. The final two examples are more complicated wormholes, the first with three asymptotic regions and the second a black hole with a single exterior but a torus behind the horizon. Entanglement in multi-boundary wormhole geometries has been considered already in \cite{Balasubramanian:2014hda}, but only between entire boundaries; the techniques here allow a relatively straightforward extension of this to subintervals.

From a more conceptual standpoint, the description of entanglement entropies obtained may be useful for understanding the status and origin of the RT and HRT prescriptions. In \cref{sec:RTHRT}, we discuss one such use, namely to show how HRT may follow from an analytic continuation of a Euclidean quantum gravity computation \cite{Lewkowycz:2013nqa} in some circumstances. This also throws up a related issue, that if geodesics computing entanglement entropy are to be understood as coming from such a Euclidean path integral, there are circumstances where there may be no geometric realisation of a relevant Euclidean geodesic in the Lorentzian spacetime. This leads to a very precise and specific notion of complexified (from the Lorentzian standpoint) entangling surface. We discuss how this may occur, and use the Euclidean version of the technology developed earlier to show in a simple example that the complex geodesics we describe are subdominant.

We conclude with a discussion of the results, and indicate some possible directions of future study. These include studies of entanglement in a variety of interesting geometries, as well as some avenues to better understand the underlying origin and properties of RT and HRT.


\section{Lengths of geodesics in quotients of $AdS_3$}

This section constructs the procedure for calculating entanglement entropy in constant negative curvature geometries. The approach we use relies on a description of $AdS_3$ as the $SL(2,\RR)$ group manifold, in which the isometries and boundary have a particularly nice algebraic structure, lending itself to construction of quotients and description of geodesics. We begin by reviewing this presentation of $AdS_3$, before describing the natural representation of the boundary in this picture. We then look at geodesics, and define a useful notion of regularised length of boundary-to-boundary geodesics. We then put this to work in the context of quotient spacetimes, which leads us to the simple algebraic computation of entanglement entropies.

There is also an analogous construction in the Euclidean setting of quotients of $H^3$, which we briefly describe in \cref{sec:H3}. We then review considerations relating Lorentzian and Euclidean computations.

\subsection{The structure of $AdS_3$}

The $AdS_3$ spacetime can be described conveniently as the group manifold of $SL(2,\RR)$. Concretely, write $\RR^{2,2}$ as

\begin{equation}
\RR^{2,2}=\left\{
p=\begin{pmatrix}
U+X & Y-V \\ Y+V & U-X
\end{pmatrix}\right\},\quad ds^2=-L_{AdS}^2\det(dp),
\end{equation}
and then $AdS_3$ is the embedded submanifold given by the one sheeted hyperboloid $\det(p)=1$. Herein, we will choose units to set $L_{AdS}=1$. Parametrising the hyperboloid by coordinates $(t,r,\phi)$, we recover the more familiar metric
\begin{equation}\label{eq:rtphi}
\left. \begin{matrix}
X= r\cos\phi, &  U= \sqrt{1+r^2}\cos t\\
Y=r\sin\phi, & V= \sqrt{1+r^2}\sin t
\end{matrix}\right\}\Rightarrow
 ds^2=-(1+r^2)dt^2+\frac{dr^2}{1+r^2}+r^2 d\phi^2
\end{equation}
with both $\phi$ and $t$ periodic\footnote{Usually, it is necessary to take the universal cover by unwrapping the time circle. We won't need to do this, since the quotient construction (save for the case of rotating BTZ) will remove the closed timelike curves.} in $2\pi$. 

The connected part of the isometry group is $SO(2,2)\cong SL(2,\RR)\times SL(2,\RR)/\ZZ_2$, which acts as $p\mapsto g_L p g_R^t$, for $g_L$ and $g_R$ in $SL(2,\RR)$, with the equivalence $(g_L,g_R)\sim(-g_L,-g_R)$. In addition to this there are two distinct discrete $\ZZ_2$ symmetries, a spatial reflection and a time reversal, the latter acting by transposition $p\mapsto p^t$.

Of particular importance is the diagonal subgroup $g_L=g_R$. This action fixes symmetric matrices, of which there are two disconnected components, being the static slices $t=0,\pi$ (distinguished by the sign of the trace). This diagonal $PSL(2,\RR)$ constitutes the set of isometries commuting with time reversal. We may identify the $t=0$ slice with the upper half plane model of the hyperbolic plane, by
\begin{equation}
p=\frac{1}{\Im(z)}\begin{pmatrix}
|z|^2 & \Re(z) \\
\Re(z) & 1
\end{pmatrix},
\end{equation}
where $z$ is in the upper half plane and $\Re(z),\Im(z)>0$ denote its real and imaginary parts. Then these isometries act on the half plane by fractional linear transformations $z\mapsto \frac{az+b}{cz+d}$ with real coefficients.

\begin{SCfigure}[2]\label{fig:Penrose}
\centering
\caption{The Penrose diagram of $AdS_3$, or equivalently $SL(2,\RR)$, formed from the rotation around the central vertical line, and identifying top ($t=\pi$) and bottom ($t=-\pi$). The boundary is at the bold lines to the left and right. The identity element is in the centre. The hyperbolic transformations are in regions \textit{\textbf{I}} ($\Tr>2$) and \textit{\textbf{II}} ($\Tr<-2$), the elliptic transformations in regions \textit{\textbf{III}}  and \textit{\textbf{IV}} ($-2<\Tr<2$), and the parabolic transformations ($\Tr=\pm2$) on the lightcones between them.
For the diagonal subgroup, hyperbolic, parabolic and elliptic classes are distinguished by having two, one or no fixed points on the boundary of the $t=0$ Poincar\'e disc respectively. Elliptic isometries also have a single fixed point in the interior of the disc.\\
Dotted lines show geodesics.}
\includegraphics[width=0.3\textwidth]{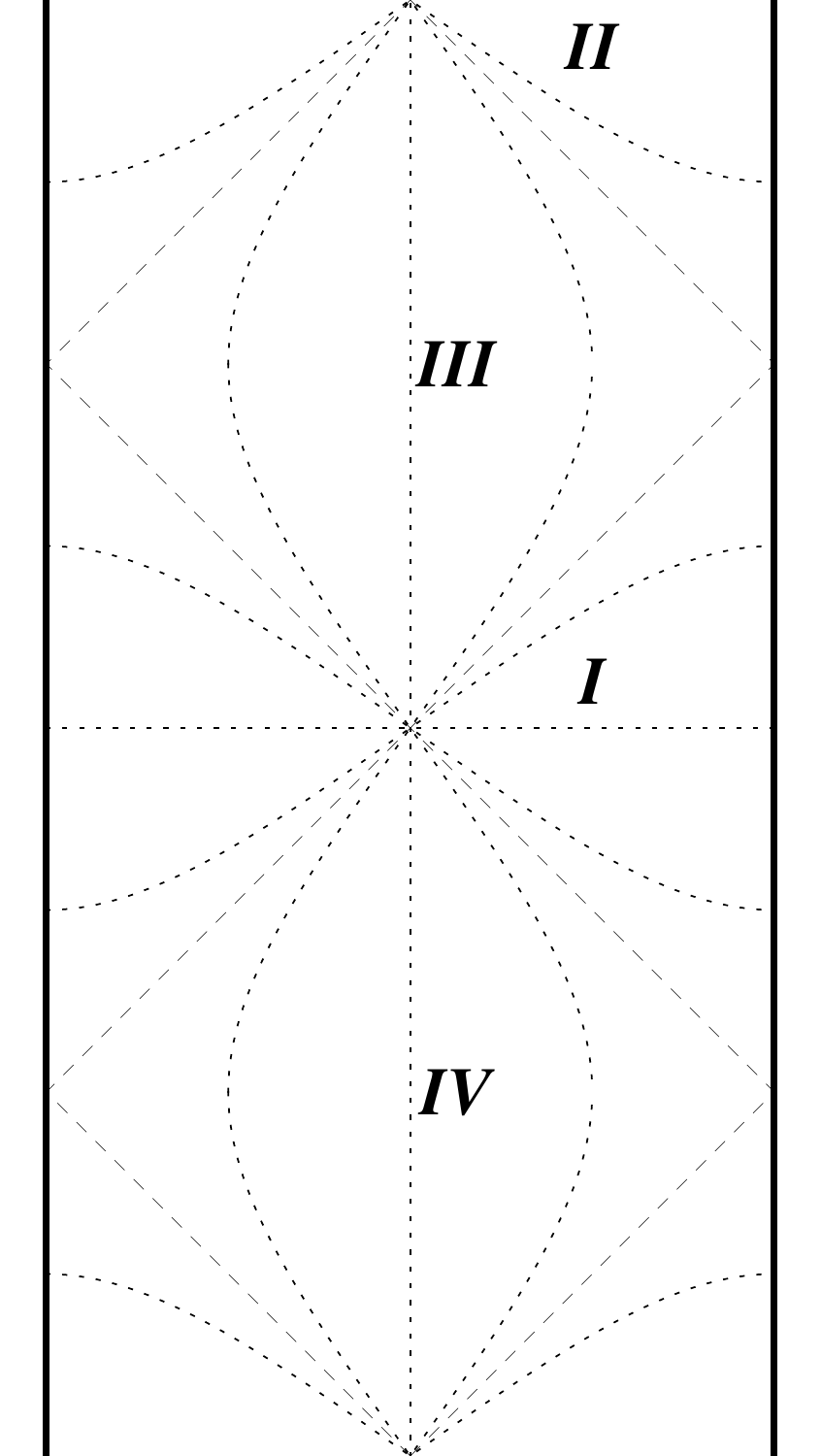}
\end{SCfigure}

The Penrose diagram in \cref{fig:Penrose} is useful to describe both the spacetime itself, as well as the elements of $SL(2,\RR)$ making up the isometries. These are divided into hyperbolic, parabolic and elliptic, depending on the trace, as described in the caption of the figure.

There is also a very natural description of the asymptotic boundary in this picture, represented by the entries of the matrix becoming large. By pulling out a factor which blows up, chosen to leave a piece with a finite limit, the points approached on the boundary can be represented by the space of singular real $2\times2$ matrices modulo overall rescaling with an arbitrary positive constant. Any such matrix can alternatively be written as $p=\vec{v}\vec{u}^t$ for some nonzero two-dimensional vectors $\vec{u},\vec{v}$. One choice of gauge fixes the overall scaling by choosing the vectors $\vec{v}$ and $\vec{u}$ to be unit, which leaves only a $\ZZ_2$ ambiguity of changing the signs of both vectors. This makes the boundary space a torus $S^1\times S^1/\ZZ_2$, and the $t=0$ slice is given by the circle $\vec{u}=\vec{v}$.

In terms of the coordinates, this corresponds to taking $r\to \infty$ at fixed $t,\phi$. With null coordinates $t=v+u$, $\phi=v-u$, the resulting matrix is
\begin{equation}
p\propto\begin{pmatrix}\cos v \\ \sin v\end{pmatrix}\begin{pmatrix}\cos u & -\sin u\end{pmatrix}
\end{equation}
and the $2\pi$ periodicities of $u,v$, along with the $\ZZ_2$ of flipping signs $(u,v)\sim(u+\pi,v+\pi)$, reproduces the $2\pi$ periodicities of $t,\phi$. From this form it is clear how the $SL(2,\RR)\times SL(2,\RR)/\ZZ_2$ acts on the boundary, with each $SL(2,\RR)$ factor acting separately as projective transformations on left- or right-moving null coordinates. Fixed points on the boundary of an isometry $(g_L,g_R)$ are points where $\vec{u}$ and $\vec{v}$ are eigenvectors of $g_L$ and $g_R$ respectively, with eigenvalues of the same sign.

The choice of positive scaling coincides precisely with a conformal factor in the boundary metric, with the unit vector choice above corresponding to the usual flat metric on the cylinder. We will review this in more detail after discussing the geodesic lengths, to which we now turn.

\subsection{Lengths of geodesics}

One advantage of this group manifold description is that the length of a geodesic between two points has a simple form. There are three cases, understood easily from the Penrose diagram \cref{fig:Penrose}.
 We are mostly interested in spacelike separated points $p,q\in SL(2,\RR)\equiv AdS_3$, which means $\Tr(p^{-1}q)>2$, so if $p$ is at the identity, $q$ is in region \textit{\textbf{I}}. There is a unique spacelike geodesic between two such points, with length
\begin{equation}
\length(p,q)=\frac12\cosh^{-1}\left(\frac{\Tr(p^{-1}q)^2}{2}-1\right)=\cosh^{-1}\left(\frac{\Tr(p^{-1}q)}{2}\right).
\end{equation}
This can be proven by using isometries to first bring $p$ to the identity, and then using the residual $SL(2,\RR)/\ZZ_2$ (with $g_R^t=g_L^{-1}$) to bring $q$ to diagonal form so the length calculation is simple. The trace requirement ensures that this is possible, since it implies that $q$ has two distinct real positive eigenvalues $e^\length$ and $e^{-\length}$. Since $\Tr(p^{-1}q)$ is invariant under the action of the isometries, we recover the stated formula in general.

For completeness, we note that the proper time along geodesics connecting timelike separated points $p$ and $q$ with $|\Tr(p^{-1}q)|<2$, (second point in region \textit{\textbf{III}} or \textit{\textbf{IV}}) is
\begin{equation}
\tau(p,q)=\cos^{-1}\left(\frac{\Tr(p^{-1}q)}{2}\right)
\end{equation}
and the choice of the branch of $\cos^{-1}$ allows for cycling round the periodic time direction any number of times. Finally, null separated points have $\Tr(p^{-1}q)=2$, and if $\Tr(p^{-1}q)\leq-2$ there is no geodesic between $p$ and $q$ (with the second point in region \textit{\textbf{II}}). The exception is the case $p=-q$ when all timelike geodesics through $p$ also pass through $q$, after proper time $\pi$.

This can be usefully applied to define a regularised length of geodesics between points on the boundary. For any boundary point, we may choose a curve $p(\rho)$ in $SL(2,\RR)$ such that at least one entry tends to infinity as $\rho\to\infty$, and $p(\rho)/\rho$ has a finite but nonzero limit $p_\partial$. This curve approaching the boundary can of course be chosen in many ways, and the resulting $p_\partial$ will differ by a positive constant of proportionality; the choice will amount to picking a regularisation scheme.

Given two such curves approaching spacelike separated points on the boundary, the distance between the points $p(\rho)$ and $q(\rho)$ is
\begin{align}
\length(p(\rho),q(\rho)) & =\cosh^{-1}\left(\frac{\Tr(p(\rho)^{-1}q(\rho))}{2}\right)\\
&= \cosh^{-1}\left(\frac{1}{2}\Tr\left( R_\perp p(\rho)^{t} R_\perp^t q(\rho)\right)\right)\\
&= \cosh^{-1}\left(\frac{\rho^2}{2}\Tr\left(R_\perp p_\partial^{t} R_\perp^t q_\partial \right)+o(\rho^2)\right)\\
&=\log \rho^2+ \log\left(\Tr\left(R_\perp p_\partial^{t} R_\perp^t q_\partial\right)\right)+o(1)\quad\text{as } \rho\to\infty
\end{align}
where $R_\perp=\left( \begin{smallmatrix}0&-1\\ 1&0\end{smallmatrix}\right)$ is the matrix rotating by angle $\pi/2$, used to implement the inverse.

Motivated by this, we define a regularised length between the boundary points by subtracting off the divergent piece:
\begin{equation}
\length_\text{reg}(p_\partial,q_\partial)  =  \log\left(\Tr\left(R_\perp p_\partial^{t} R_\perp^t q_\partial\right)\right).
\end{equation}
Notice that this depends on the choice of singular matrix to define each point, but the boundary is defined as the set of such matrices modulo positive rescalings; this is the dependence on the choice of regularisation.

If we choose to express the boundary points as $p_\partial=\vec{u_1}\vec{v_1}^t$, and $q_\partial=\vec{u_2}\vec{v_2}^t$, then this can also be expressed very simply as 
\begin{equation}
\length_\text{reg}(p_\partial,q_\partial)=\log\left((\vec{u}_1^\perp\cdot\vec{u}_2)(\vec{v}_1^\perp\cdot\vec{v}_2)\right),
\end{equation}
where $\vec{u}^\perp=R_\perp \vec{u}$ represents the rotation of $\vec{u}$ by $\pi/2$. This could alternatively be expressed with two-dimensional cross products of $\vec{u}$s and $\vec{v}$s, or determinants.

The choice of regularisation we describe here maps very simply to the usual IR cutoff prescription, where the lengths are measured up to some fixed radius in a Fefferman-Graham expansion. This expansion depends on the choice of boundary metric, which is fixed simply by the conformal factor relating it to the flat metric on the cylinder. Indeed, one may choose $\rho$ to be the radial Fefferman-Graham coordinate with the preferred choice of boundary metric, in which case the prescription is manifestly equivalent to the usual procedure of regulating at constant $\rho$ and dropping the cutoff dependent piece. 

The flat metric comes from choosing $\vec{u}$ and $\vec{v}$ to be unit vectors as described above. To illustrate, on the static $t=0$ slice choosing $\vec{u}=\vec{v}=(\cos(\phi/2),\sin(\phi/2))^t$ gives $\length_\text{reg}=\log\sin^2(\Delta\phi/2)$, which, up to a factor of $\frac{1}{4G_N}=\frac{c}{6}$ and the constant non-universal piece, reproduces the universal 2D CFT entanglement entropy on a circle in vacuum as expected.

When we move to discussing quotients, it will be necessary to make another choice of boundary metric, in particular one that is invariant under the action of the quotient group. If this is related to the flat metric as
\begin{equation}
ds^2 = \Omega^2 ds^2_{\text{flat}}
\end{equation}
then the representative matrix on the boundary is just a factor of $\Omega$ larger than one built from unit vectors.

As a simple example of an alternative regularization on the $t=0$ slice, natural in Poincar\'e patch or planar coordinates, choose $\vec{u}=\vec{v}=(x,1)^t$. This differs from the vacuum normalization by a conformal factor of $\Omega=(1+x^2)$, which is just right to give the metric of Minkowski space when extended to the diamond $-\pi<t\pm \phi<\pi$. The regularised length gives the entanglement entropy result $\frac{c}{3}\log\Delta x$, being the universal CFT result on a line in vacuum.

\subsection{Quotients}

Any constant negative curvature geometry can be obtained from a quotient of some subset of $AdS_3$ by a discrete group of isometries\footnote{Some, such as planar BTZ, must be obtained as a limit of such quotient geometries.}. The technology we develop for computing regularised lengths of geodesics relies of having such a description for the spacetime of interest. We only briefly outline the aspects of the construction we need; for the full details see \cite{Aminneborg:1998si} for the general case, and \cite{Aminneborg:1997pz,Brill:1995jv,Brill:1998pr} for more on the special non-rotating case. An extension to allow orientation reversing isometries is relatively straightforward; we will not describe the details here, though an example will be given in section \cref{sec:RP2}.

Suppose we have a discrete group $\Gamma\subseteq SL(2,\RR)\times SL(2,\RR)/\ZZ_2$ of hyperbolic isometries. The hyperbolic condition here means that the trace of both $SL(2,\RR)$ elements is greater than two, and ensures that the spacetime is free of conical singularities, closed timelike curves, or other singular features\footnote{Some other special classes of isometries including parabolic elements are important, for example for the extremal BTZ black hole (see \cite{Aminneborg:1998si,Banados:1992gq}), but only minor details change in what follows.}. We can form a spacetime as the quotient $\widehat{AdS}/\Gamma$ of a particular covering space $\widehat{AdS}$ by the isometries. The covering space is formed from $AdS_3$ by removing all points to the future and past of any fixed point of an isometry lying on the boundary. In particular, the covering space on the boundary will be a (generically infinite) collection of diamonds with corners at the fixed points of some $\gamma\in\Gamma$. The resulting bulk covering space is simply connected, so the fundamental group of the resulting geometry can be identified with $\Gamma$.

It is easiest to understand the construction in the special `non-rotating' case of $\Gamma$ being a subgroup of the diagonal $PSL(2,\RR)$, acting on the $t=0$ slice, modelled as the upper half plane, by fractional linear transformations. The quotient of the upper half plane obtained, which is a time-reflection invariant slice of the quotient geometry, will be a surface with some number of boundaries and handles. This can be thought of as any number of black hole exteriors joined behind the horizon with an arbitrary topology. See \cref{sec:wormhole3,sec:Torus} for examples.

A geodesic in the quotient is represented by a set of geodesics in the covering space, related by elements of $\Gamma$. Between any two spacelike separated boundary points $p$ and $q$ in $\widehat{AdS}/\Gamma$, there will be many geodesics, in one-to-one correspondence with elements of $\Gamma$. To see this, pick some representatives $\hat{p}$ of $p$ and $\hat{q}$ of $q$ in covering space. Each geodesic in the quotient will have a unique lift ending on $\hat{p}$. The other end of this lifted geodesic may be at any of the points $\gamma\hat{q}$, with $\gamma\in\Gamma$, and these are all distinct since fixed points have been removed from the covering space. Topologically, this correspondence between geodesics and elements of $\Gamma$ results from the fact that each homotopy class of curves has a unique geodesic representative.

There is a subtlety here, that the lifted geodesic could leave the covering space $\widehat{AdS}$, and hence not be an admissible geodesic in the quotient, but we show that this is impossible. This relies on a property of any spacelike geodesic in $AdS$. Begin by constructing the pair of diamonds of boundary points spacelike separated from the endpoints of the geodesic. Now take the `causal wedges' (see \cite{Hubeny:2012wa}) in the bulk formed by the union of causal curves beginning and ending in one of these diamonds. The causal wedges are bounded by null surfaces to the past and future, and these intersect precisely at the geodesic. The consequence of this is that the past of the geodesic on the boundary is exactly the past of its endpoints, being the past of the pair of diamonds. Another way of stating this is that the geodesics satisfy causality in the sense of \cite{Headrick:2014cta}, but only marginally. Now if there is a fixed point of $\Gamma$ on the boundary in the past of the geodesic, then it is also in the past of at least one of the endpoints. But in that case, the endpoint is not in the covering space, and does not represent a point on the boundary of the quotient geometry. Identical remarks apply with `past' replaced by `future', so no point of the geodesic leaves $\widehat{AdS}$.



Now, we wish to compare the lengths of the various geodesics between two boundary points in the quotient. The lift of these geodesics to covering space will connect different endpoints, related by elements of $\Gamma$. To make sure we are regulating consistently, we must ensure that the points at which we impose the infrared cutoff in different regions are images of one another under the appropriate element of $\Gamma$. The result is that the singular matrices representing points on the boundary may not be chosen independently, but must be invariant under the action of $\Gamma$: we may only choose a regularisation for a single lift of each point, and this will determine all its images' regularisations. This choice of singular matrices can be considered as a choice of boundary metric, which must be invariant under the action of $\Gamma$ on the boundary.

So for each pair of points in $\partial\widehat{AdS}/\Gamma$, having chosen singular matrices $p,q$ to represent a lift of each,  the consistent regularised lengths of geodesics between the points are given by
\begin{equation}\label{eq:openLength}
\ell_\text{reg}(p,q,\gamma)
=\log\left(\Tr\left(R_\perp p^{t} R_\perp^t g_L q g_R^t \right)\right)
=\log\left((\vec{u}_1^\perp\cdot g_L\vec{u}_2)(\vec{v}_1^\perp\cdot g_R\vec{v}_2)\right)
\end{equation}
where $\gamma=(g_L,g_R)\in \Gamma\subseteq SL(2,\RR)\times SL(2,\RR)/\ZZ_2$ and $p=\vec{u_1}\vec{v_1}^t$, $q=\vec{u_2}\vec{v_2}^t$.

In any quotient, the asymptotic regions approaching each boundary component are isometric to the external region of the BTZ black hole up to the horizon. Hence, a natural way to choose the regularisation is to reproduce the BTZ result for lengths of geodesics remaining outside the horizons, picking flat cylinder metrics on each boundary component. Locally, near each boundary, there will be two Killing vectors generating translations in space and time (though not extendible consistently into the whole interior). With this regularisation choice, matrices representing different points in the same asymptotic boundary will be related by the flow of these local symmetries. This leaves only one overall constant to be determined, fixed by matching to some standard normalisation for lengths between points at small separation, say $S\sim\frac{c}{3}\log l$ for small intervals of length $l$.

For computing entanglement entropy holographically, apart from geodesics ending on the boundary it is also required to be able to compute lengths of closed spacelike geodesics. 
A closed geodesic in the quotient again lifts to a set of geodesics in the covering space $\widehat{AdS}$, each lift connecting a pair of fixed points on the boundary of some $\gamma\in\Gamma$. This $\gamma$ is the element which takes any point on the curve in covering space to a corresponding point obtained by following the geodesic for a single cycle. A different lift of the closed geodesic will connect fixed points of some $\gamma'$ conjugate to $\gamma$ in $\Gamma$. Hence, the closed geodesics are in one to one correspondence with conjugacy classes of $\Gamma$. From a topological point of view, this says that each free homotopy class of loops\footnote{The standard definition of homotopies in the definition of the fundamental group requires beginning and ending loops on some fixed basepoint, necessary to give $\pi_1$ its group structure. Free homotopy here means homotopies of loops without the requirement to fix the basepoint. The equivalence classes under this larger set of homotopies are conjugacy classes of the fundamental group, conjugation corresponding to taking different paths linking a loop back to the basepoint.} has exactly one geodesic representative.

This also gives us a simple way to compute the length of the closed geodesic. We can use the isometries of $AdS$ to send the fixed points of $\gamma$ to $t=\phi=0$ and $t=0,\phi=\pi$, and hence the geodesic to the straight line across the $t=0$ slice (being the diagonal matrices). It is then straightforward to compute
\begin{equation}\label{eq:closedLength}
\ell=\cosh^{-1}\left(\frac{\Tr g_L}{2}\right)+\cosh^{-1}\left(\frac{\Tr g_R}{2}\right),
\end{equation}
where $\gamma=(g_L,g_R)$ for $g_L$ and $g_R$ in $SL(2,\RR)$, with traces greater than 2 as required for the transformation to be hyperbolic.
%

\subsection{Homology}

In this description it is very easy to compute the homology classes of curves in order to algorithmically decide which combinations of open and closed geodesics are topologically admissible. We will take the criterion to be that the geodesics in the bulk, together with the intervals on the boundary whose entanglement we compute, when given appropriate orientations, should have trivial homology\footnote{This is in fact not quite the right thing if the spacetime is nonorientable. The required modification will be discussed in \cref{sec:RP2}.}. In other terms, there must be a codimension-one oriented submanifold whose boundary is the interval along with the geodesics. This is essentially the prescription used to prove that known quantum mechanical properties of entanglement entropy, such as strong subadditivity, follow from the holographic prescription \cite{Headrick:2007km,Wall:2012uf}. In the  case of time-reflection symmetric geometries at the moment of time symmetry, where a Euclidean quantum gravity approach is available, this form of the homology constraint can be proven \cite{Haehl:2014zoa} under certain assumptions about the holographic replica trick \cite{Lewkowycz:2013nqa}.

As already commented upon, the fundamental group of the quotient space is isomorphic to the quotient group $\Gamma$ itself. The first homology group $H_1$ is the abelianisation of the fundamental group, so homology is obtained from homotopy simply by counting the number of each generator contained in any element $\gamma\in \Gamma$.

For closed geodesics, this is entirely straightforward: closed curves are associated with conjugacy classes of $\Gamma$, any representative of a conjugacy class will have the same nett number of each generator, and so any representative will do to compute the homology.

For open geodesics, it is equally straightforward if the representatives of boundary points are chosen appropriately. Any interval on the boundary of the quotient lifts to a collection of intervals on the boundary of covering space, and if the representatives of the endpoints are chosen from ends of the same component of the lifted interval, then the element $\gamma$ associated with a geodesic will correspond to the homology class of the geodesic along with the interval. This means that the interval will be homotopic to the geodesic associated with the identity in $\Gamma$. This will happen automatically if the representatives are obtained by translating with local Killing vectors in each component.

If there are geodesics between different asymptotic boundaries, and the endpoints of the intervals are chosen as above, this can be straightforwardly extended by combining the abelianisations of all elements of $\gamma$ from the geodesics, taking care with signs\footnote{An individual geodesic going between different asymptotic regions does not have a meaningful homology class on its own, since without other bulk geodesics it can never form a closed loop. Only combinations that can be summed to make cycles have any invariant meaning.}.

\subsection{The story in $H^3$}\label{sec:H3}

There is an analogous picture in the Euclidean hyperbolic 3-space, by taking Hermitian (instead of real) $2\times2$ matrices of unit determinant. In fact, this gives two disconnected copies of $H^3$, which can be distinguished as positive and negative trace components.
\begin{equation}
	H^3=\{M\in SL(2,\CC)|M=M^\dag, \Tr M>0\}
\end{equation}
Explicitly, this can be seen as the top sheet of the two-sheeted hyperboloid embedded in $\RR^{3,1}$ written as
\begin{equation}
\RR^{3,1}=\left\{
p=\left(\begin{matrix}
T+Z & X-iY \\ X+iY & T-Z
\end{matrix}\right)\right\},\quad ds^2=-\det(dp),
\end{equation}
given by $\det(p)=1$. The normal ball model coordinates are
\begin{equation}
\left. \begin{matrix}
(X,Y,Z)= \frac{2}{1-r^2}(x,y,z)\\
T=\frac{1+r^2}{1-r^2}
\end{matrix}\right\}\Rightarrow
 ds^2=\frac{4(dx^2+dy^2+dz^2)}{(1-r^2)^2}
\end{equation}
where $r^2=x^2+y^2+z^2<1$.

Now the connected part of the isometry group is given by $SO(2,1)\cong PSL(2,\CC)$, acting by $p\mapsto gpg^\dag$. There is one orientation reversing isometry, acting by transposition. This fixes an equatorial copy of the hyperbolic plane at $Y=0$, where the matrices are real symmetric, matching the static $t=0$ slice of $AdS_3$ precisely, and it is fixed by the obvious $PSL(2,\RR)\subseteq PSL(2,\CC)$. This matching between the diagonal symmetry group in the Lorentzian case and real part of the symmetry group in the Euclidean case, along with identification of the static slice, will play an important r\^ole in certain contexts, as reviewed in \cref{sec:EuclidLorentz}.

Once again, the boundary is approached when any entry of the matrix becomes large, so is represented by singular $2\times2$ Hermitian matrices of positive trace, modulo real rescaling. These can be written as $\vec{u} \vec{u}^\dagger$ for some $\vec{u}\in \CC^2-\{0\}$, with an equivalence under scaling $\vec{u}$ by a nonzero complex number. In other words, the boundary is just the Riemann sphere $\CC\PP^1$. One particular canonical choice of $\vec{u}=(z,1)$ can be made, for $z\in \CC\cup\{\infty\}$, in terms of which the isometry group then simply acts on the boundary by fractional linear transformations in the usual way.

Lengths of geodesics are again given by the formula
\begin{equation}
\length(p,q)=\cosh^{-1}\left(\frac{\Tr(p^{-1}q)}{2}\right)
\end{equation}
and the computation for regularised length between boundary points remains essentially unchanged, giving
\begin{align}
\length_\text{reg}(p,q)  &=  \log\left(\Tr\left(R_\perp p^{t} R_\perp^t q\right)\right) \\
&= \log\left|\vec{u}_\perp\cdot \vec{v}\right|^2
\end{align}
where $p=\vec{u}\vec{u}^\dag$, $q=\vec{v}\vec{v}^\dag$ and $\vec{u}_\perp=R_\perp \vec{u}$ as before.

Hyperbolic manifolds can be obtained by taking a quotient $H^3/\Gamma$ with a discrete subgroup $\Gamma$ of $PSL(2,\CC)$, known as a Kleinian group, consisting of loxodromic or parabolic elements (excepting the identity) so that the quotient group acts without fixed points in the bulk. This is simpler than the Lorentzian case, since no points of the bulk need to be removed from the covering space. The boundary of the space will be a quotient by $\Gamma$ of a set of points on which $\Gamma$ acts `nicely', known as the domain of discontinuity (which may be empty, giving a `closed universe') and will result in a Riemann surface or orbifold.

Two special cases are worth mentioning here. Handlebody geometries, which loosely speaking are obtained from `filling in' a closed Riemann surface of genus $g$, are obtained from Schottky groups with $g$ generators, as discussed extensively in \cite{Krasnov:2000zq}. There are more ways to obtain a bulk with a given Riemann surface as its boundary, though there is a conjecture that the handlebodies are the dominant geometries holographically \cite{Yin:2007at}. Secondly, the groups which fix the equatorial slice are discrete subgroups of $PSL(2,\RR)$, known as Fuchsian groups, and these offer the most obvious mapping between Lorentzian and Euclidean spacetimes. These two classes overlap, so a group may be both Fuchsian and Schottky.

The lengths of geodesics in quotients are found in much the same manner as the Lorentzian case. The relevant formulae for geodesics joining boundary points represented by singular matrices $p=\vec{u}\vec{u}^\dag,q=\vec{v}\vec{v}^\dag$ is
\begin{equation}\label{eq:openLengthEuclid}
\ell_\text{reg}(p,q,\gamma)
=\log\left(\Tr\left(R_\perp p^{t} R_\perp^t g q g^\dag \right)\right)
=\log|\vec{u}_\perp\cdot g \vec{v}|^2
\end{equation}
for $g\in\Gamma$. The lengths of closed geodesics are given by
\begin{equation}\label{eq:closedLengthEuclid}
l=\cosh^{-1}\left[\left|\frac{Tr g}{2}\right|^2+\left|\left(\frac{Tr g}{2}\right)^2-1\right|\right]
\end{equation}
where $g$ is a representative of the conjugacy class of $PSL(2,\CC)$ associated to the curve. Note that in the real $PSL(2,\RR)$ case, these formulae all match the Lorentzian formulae in the diagonal $PSL(2,\RR)$ case.

\subsection{Matching between Euclidean and Lorentzian descriptions}\label{sec:EuclidLorentz}

The Euclidean geometries offer a natural way to describe the CFT state dual to a given quotient geometry, at least if the quotient group $\Gamma$ is in the diagonal $PSL(2,\RR)$ \cite{Skenderis:2009ju,Krasnov:2000zq}, generalising the duality between the eternal black hole and the thermofield double state \cite{Maldacena:2001kr}.

 In this case there is a moment of time-reflection symmetry, which automatically has zero extrinsic curvature, so can also be described as a surface in a Euclidean geometry, which can be obtained by the quotient of $H^3$ by the same group $\Gamma$, now in the real $PSL(2,\RR)$ subgroup. The state on the $t=0$ slice can now be defined by a Hartle-Hawking procedure as being prepared by a path integral over the bottom half of the Euclidean spacetime.

The interpretation of the CFT state is the boundary analogue of this construction: a path integral over a Riemann surface with boundaries defines a state on those boundaries, as the wave functional evaluated on the field configuration given by the boundary conditions. The partition function is obtained from gluing a pair of these surfaces together at their boundaries, to give the Schottky double.

However, from any given Riemann surface, there will in general be many different bulk geometries with that boundary, and the partition function may be computed from any one of them depending on which is the dominant saddle point with smallest Euclidean action. As the moduli vary, the dominant saddle point will change, and the Euclidean bulk description will change. These transitions are a generalisation of the Hawking-Page phase transition. Assuming that the bulk does not spontaneously break the time reflection symmetry, the appropriate Lorentzian geometry can then be obtained by evolving from the time-reflection invariant slice of the dominant Euclidean saddle point. This may be disconnected, which results in a Lorentzian spacetime consisting of several disjoint components, though they are all connected in the Euclidean section. See also \cite{Balasubramanian:2014hda} for a more detailed discussion.

\section{Examples}\label{sec:examples}

\subsection{BTZ}

The simplest example of a quotient of $AdS$ is the BTZ black hole \cite{Banados:1992wn,Banados:1992gq}, generated by a single hyperbolic element $\gamma$ of $SL(2,\RR)\times SL(2,\RR)/\ZZ_2$. By conjugation, this can be chosen to be the result of exponentiating the $\mathfrak{sl}(2,\RR)\oplus\mathfrak{sl}(2,\RR)$ Lie algebra elements (or equivalently Killing vectors)
\begin{equation}
\xi_L=
\frac{r_+-r_-}{2}
\begin{pmatrix}
1 & 0 \\ 0 & -1
\end{pmatrix},
\quad
\xi_R=
\frac{r_++r_-}{2}
\begin{pmatrix}
1 & 0 \\ 0 & -1
\end{pmatrix}
\end{equation}
with $r_+\geq r_-\geq0$, so $\gamma=(g_L,g_R)=(\exp(2\pi\xi_L),\exp(2\pi\xi_R))$. The cases $r_+=r_-$ and $r_-=0$ correspond to extremal and static cases respectively.

This isometry fixes the two points $t=0,\phi=0$ and $t=0,\phi=\pi$ on the boundary, and so the boundary covering space is the pair of diamonds spacelike separated from both these points. As a result, the spacetime has two asymptotic boundaries, with the field theory interpretation of the two noninteracting but entangled halves of a thermofield double state.

This spacetime has two independent Killing vectors, generated by the $\mathfrak{sl}(2,\RR)$ matrices commuting with $\xi_L$ and $\xi_R$, acting on the left and right respectively. These can be chosen as $\xi=(\xi_L,\xi_R)$ itself, generating spatial translations on the boundaries, and the orthogonal combination $\eta=(\xi_L,-\xi_R)$ generating time translations, acting in opposite directions in the two asymptotic regions.

The $SL(2,\RR)$ description of the spacetime can be related to the more usual coordinates $(t,r,\phi)$ in one exterior region $r>\rp$ by
\begin{equation}
p=\frac{1}{\sqrt{\rp^2-\rm^2}}
\begin{pmatrix}
 e^{-r_- t+r_+ \phi}\sqrt{r^2-r_-^2} &  \quad e^{-r_+ t+r_- \phi}\sqrt{r^2-r_+^2} \\
  e^{r_+ t-r_- \phi}\sqrt{r^2-r_+^2} &  \quad e^{r_- t - r_+ \phi}\sqrt{r^2-r_-^2} \\
\end{pmatrix}.
\end{equation}
Acting with $\gamma$ simply maps $\phi$ to $\phi+2\pi$, so implements the periodicity of the angular direction. The metric is then
\begin{equation}
ds^2=-f(r) dt^2 + \frac{dr^2}{f(r)} + 
 r^2 \left(d\phi - \frac{r_+ r_-}{r^2}dt\right)^2,\quad \text{where } f(r)=\frac{(r^2-r_+^2)(r^2-r_-^2)}{r^2}.
\end{equation}
In terms of the inner and outer horizon radii, the physical parameters of the black hole are its mass $M=r_+^2+r_-^2$ and angular momentum $J=2r_+r_-$.

There is one closed geodesic for each element $\gamma^n$ of $\Gamma$, according to the correspondence with conjugacy classes of $\Gamma$. These are geodesics wrapping the bifurcation circle $n$ times, and have length $2\pi|n|r_+$, as can be checked from equation \eqref{eq:closedLength}.

The length calculations require choosing a singular matrix, or a pair of vectors, to represent one point on each boundary, normalised in some convenient way. A simple set of choices is at $t=\phi=0$, with
\begin{equation}
\vec{u}_1=\vec{v}_1=(r_+^2-r_-^2)^{-1/4}\begin{pmatrix}1\\1\end{pmatrix},\quad
\vec{u}_2=\vec{v}_2=(r_+^2-r_-^2)^{-1/4}\begin{pmatrix}-1\\1\end{pmatrix}
\end{equation}
with the subscripts labelling the region. The factor is chosen for a convenient standard normalisation. The regularised lengths are now extremely simple to compute, from equation~\eqref{eq:openLength}, and by translating the points to the required locations with the Killing vectors $\xi$ and $\eta$. Since these are Killing vectors, this translation need only be done to one endpoint, the other being placed at $t=\phi=0$. Furthermore, the inclusion of group elements of $\Gamma$ labelling different geodesics corresponds to adding multiples of $2\pi$ to $\phi$ to wrap round the spatial circle, so the lengths are all obtained from computing
\begin{equation}
\ell
=\log\left((\vec{u}_i^\perp\cdot \exp(\phi\xi_L+t\eta_L)\vec{u}_j)
(\vec{v}_i^\perp\cdot \exp(\phi\xi_R+t\eta_R)\vec{v}_j)\right).
\end{equation}
 The result between points on the same exterior region is
\begin{equation}
\ell=\log \left[\frac{4}{r_+^2-r_-^2}\sinh\left(\frac{r_++r_-}{2}(\phi-t)\right) \sinh\left(\frac{r_+-r_-}{2}(\phi+t)\right]
\right)
\end{equation}
and for geodesics between opposite regions it is
\begin{equation}
\ell=\log \left[\frac{4}{r_+^2-r_-^2}\cosh\left(\frac{r_+-r_-}{2}(\phi+t)\right) \cosh\left(\frac{r_++r_-}{2}(\phi-t)\right)
\right]
\end{equation}
where it should be borne in mind that the time evolution here is by the `$H_L-H_R$' Killing field which acts in opposite directions on the two boundaries. The first of these results reproduces the answer obtained with the original HRT proposal \cite{Hubeny:2007xt}, slightly generalised to allow for endpoints of the interval to lie at different times. The second can in fact be found from the first by analytically continuing in $t$, and the result is given for example in \cite{Iizuka:2014wfa}.

With these expressions we could go on to study entanglement entropies of intervals, mutual information, including entanglement between the two halves of the thermofield double, and so forth. This has already been considered in detail in special cases, for example \cite{Hartman:2013qma, Morrison:2012iz}, so we will not dwell on this but instead move on to our next example.

\subsection{$\RR \PP^2$ geon}\label{sec:RP2}

We turn now to a simple example of a quotient including orientation reversing isometries, the $\RR\PP^2$ geon, first discussed in relation to holography in \cite{Louko:1998hc}.

Spatial reflection in $AdS_3$ is implemented by conjugation with a traceless matrix of determinant $-1$, hence with eigenvalues $\pm 1$ and squaring to the identity, for example $\left(\begin{smallmatrix}1&0\\0&-1\end{smallmatrix}\right)$. Combining this with a hyperbolic element gives an isometry
\begin{equation}\label{eq:RP2quotient}
\gamma: p\mapsto g p g^t,\quad 
g=\begin{pmatrix} \exp\left(\frac{\pi r_+}{2}\right)&0\\0&-\exp\left(\frac{-\pi r_+}{2}\right) \end{pmatrix}.
\end{equation}
The geon is obtained by taking the quotient under the group generated by $\gamma$, isomorphic to $\ZZ$.

The element $\gamma^2$ is an orientation preserving isometry, and the quotient under the subgroup of the even elements generated by it gives the nonrotating case of the BTZ black hole. Adding in the odd elements corresponds to taking the quotient of the black hole under the involution which swaps the two asymptotic regions, and also rotates half way round the spatial circle. This involution acts without fixed points, so the result is a smooth spacetime (except at the usual BTZ singularities). This leaves a geometry with a single exterior region, isometric to the black hole until the centre of the Einstein-Rosen bridge where the spatial circle has antipodal points identified, as an $\RR\PP^1$. Any time slice thus has the topology of a cylinder with the boundary at one end and a cross-cap inserted at the other, which gives a M\"obius strip, or alternatively the real projective plane $\RR\PP^2$ with a boundary, whence the name.
\begin{SCfigure}[2]\label{fig:RP2Penrose}
\centering
\includegraphics[width=0.3\textwidth]{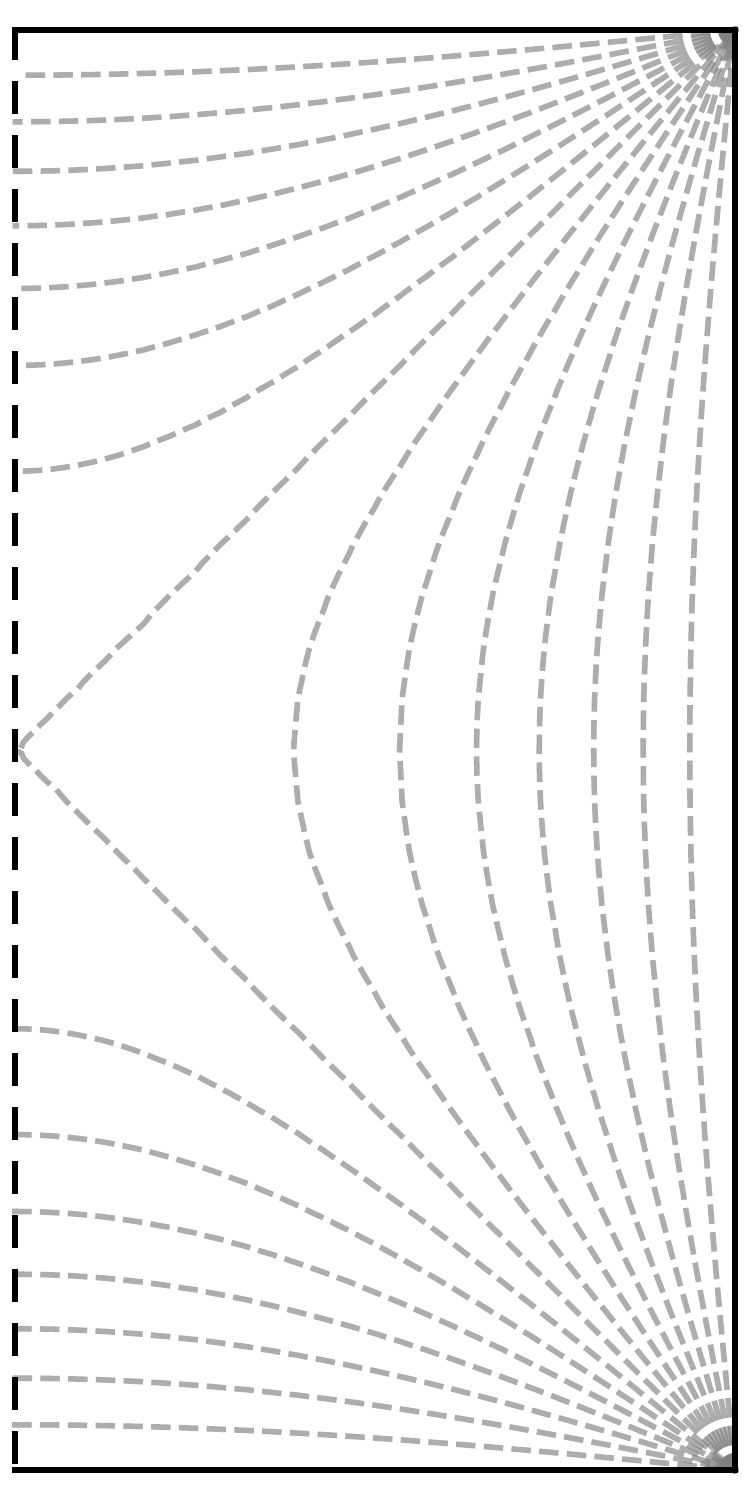}
\caption{The Penrose diagram for the $\RR\PP^2$ geon. The light dashed lines are of constant radius $r$, including the event horizon at $r=\rp$. Each point on the diagram is a circle, with antipodal points identified at the dashed line on the left. The boundary is on the right, and past and future singularities at the bottom and top.}
\end{SCfigure}

In $d+1$ dimensions, this construction can be generalised to an $\RR\PP^d$ geon by taking a quotient of a Schwarzschild or other more general spherically symmetric black hole with two exterior regions, by the involution swapping the asymptotic regions combined with the antipodal map on the sphere \cite{Giulini:1989,Friedman:1993ty}.

The $\RR\PP^2$ geon geometry has a single Killing vector field,
\begin{equation}
\xi_L=\xi_R=\frac{r_+}{2}\begin{pmatrix}1&0\\0&-1\end{pmatrix}
\end{equation}
generating spatial translations. The time translation symmetry of the BTZ black hole is broken by the reflective involution, so is not smoothly extendible into the whole spacetime, so the geon is not stationary. It still generates time translations on the boundary, and exists outside the event horizon as the local Killing vector field $\eta=(\xi_L,-\xi_R)$.

From a field theory point of view, the state at $t=0$ can be prepared by a path integral over a M\"obius strip, exactly as in the orientable case described in \cref{sec:EuclidLorentz}. It is therefore a pure state on a single copy of the CFT. From the Euclidean point of view, swapping the asymptotic regions of BTZ can be viewed as a refection in the Euclidean time, $t_E\mapsto \beta/2-t_E$, which fixes $t_E=\pm\beta/4$. Taking the quotient of the torus by this reflection along with $\phi\mapsto\phi+\pi$ gives the resulting boundary geometry of a cylinder of length $\beta/2$, with cross-caps at both ends: a Klein bottle. The partition function is given by a path integral over this Klein bottle, a surface with a single real modulus identified with $\rp$.

Unlike the torus, holographically this state has only one phase. Any phase must be obtainable from a Euclidean bulk with toroidal boundary (the oriented double cover) by taking a quotient by the appropriate involution, which restricts to the correct thing on the boundary torus. This boundary involution extends into the bulk in the black hole phase (giving the geon), and also in the `thermal AdS' phase where the spatial circle is contractible in the bulk. But in the latter case the involution acts with fixed points at $t_E=\pm \beta/4$, and the resulting Euclidean bulk is not smooth. The consequence is that there is a single phase at all moduli of the boundary, and there is no analogy to the Hawking-Page transition.

One qualitative way to think of the state, made more precise in \cite{Louko:1998hc}, is that it is very similar to a thermofield double, being made up of two halves that look thermal individually, but purify one another when taken together. But now the halves are not two copies of the CFT, but the left- and right-movers in a single copy. 

Now computation of the lengths of geodesics can be done using the methods described above essentially without modification. In particular, the equation~\eqref{eq:openLength} still applies, excepting that the matrices describing the quotient may have determinant $-1$ for orientation reversing elements.

A pair of representative vectors for the boundary is $\vec{u}=\vec{v}=(1,1)^t /\sqrt{\rp}$, and representatives at other times can be found by inserting exponentials of Killing vectors as for BTZ. There are two classes of geodesics, one staying away from the horizon, and one passing through it, which corresponds to inserting a factor of $g$. The results for lengths of geodesics connecting points at equal times, separated by an angle $\phi$, are
\begin{align}
\length_1 &= \log\left(\frac{4}{\rp^2}\sinh^2\left(\frac{\rp\phi}{2}\right)\right) \\
\length_2 &= \log\left(\frac{2}{\rp^2}\left(\cosh (\rp (\pi-\phi))+\cosh (2\rp t)\right)\right),
\end{align}
which can also be obtained directly from the appropriate results for BTZ.
Additionally, there is one simple closed geodesic, going half way round the bifurcation circle, closed due to the identifications, with length $\pi \rp$.

This state throws up an interesting question of how precisely to state the topological constraint on allowed geodesics, since if the statement is that the boundary region together with the geodesic should have trivial homology class in $H_1(\mathcal{M},\mathbb{Z})$, we arrive at a contradiction. The state is pure, so computing the entropy of the whole boundary should allow for no geodesic at all: the boundary itself should have trivial homology. But this is not the case in the given definition, since the interpolating manifold filling a constant time slice bulk is not a chain over $\mathbb{Z}$, being nonorientable. The solution\footnote{One proposed solution might be to take homology with coefficients in $\mathbb{Z}_2$. This turns out to be a necessary condition, but it is not a strong enough constraint in general.}, described in detail in \cite{Haehl:2014zoa} motivated by a holographic replica trick derivation \cite{Lewkowycz:2013nqa}, is to insist on the existence of a spacelike `two-sided' codimension-1 manifold, where two-sided means it has a globally defined continuous unit normal vector. In an orientable spacetime, this is equivalent to the homology constraint as stated above, but does not follow from such a homology theory in nonorientable spacetimes. It is this version that we will employ in what follows.

The result for the entanglement entropy of a single interval is identical to the answer in the thermal case when the interval is less than half of the boundary. The answers differ when the interval is larger than half since the state is pure, so entanglement entropy of a region equals the entanglement entropy of the complement. The length of the geodesic staying away from the horizon is always shorter than the geodesic passing through it, when combined with the closed geodesic required by homology. This can be seen directly from the expressions obtained: the difference $\length_2+\pi \rp-\length_1$ is smallest when $t=0$ and $\phi=\pi$, equalling $\rp\pi-\log\sinh^2\left(\frac{\rp \pi}{2}\right)$. This is bounded below by $\log4$, so always positive.

In fact this can be seen directly from the geometry without any calculations. On the $t=0$ slice, the geodesic passing through the centre can be joined to the closed geodesic at the bifurcation circle, but the combination is a single curve with sharp corners, and so can be reduced in length by smoothing these out. Hence there is a shorter curve, and hence a shorter geodesic, on the $t=0$ slice. For nonzero times, it is not even clear that the geodesic passing through the centre is allowable, since there is no closed geodesic to add that will allow for a spacelike interpolating surface, required in a version of the topological constraint including causality. Even allowing it, the geometric argument can be straightforwardly adapted in any case, since the central $\RR\PP^1$ is largest at the bifurcation point.

\begin{figure}
\centering
\begin{subfigure}[t]{0.45\textwidth}
\centering
\includegraphics[width=\textwidth]{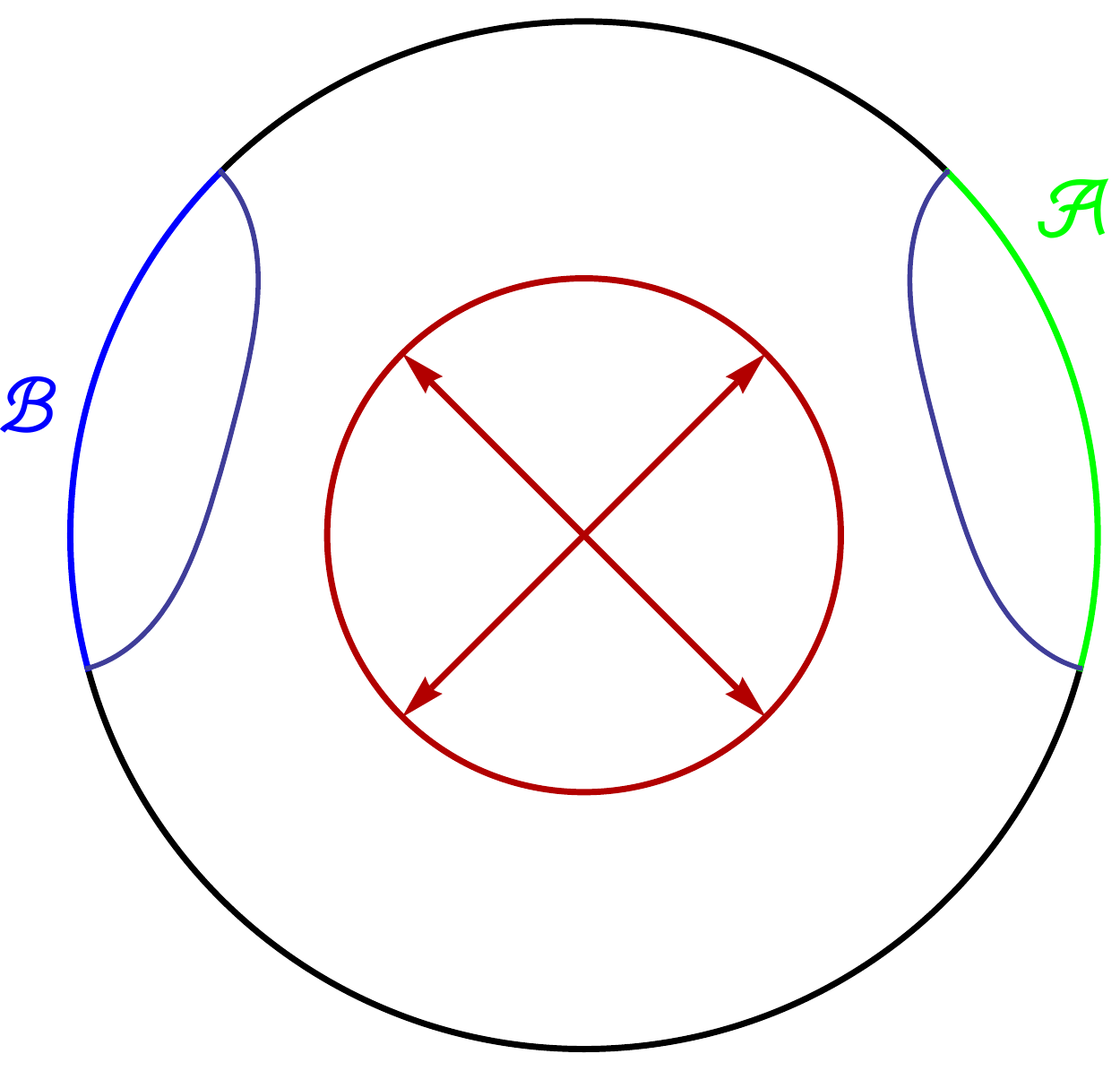}
\caption{Phase 1. Dominates for small $l$.}
\end{subfigure}
\hspace{.07\textwidth}
\begin{subfigure}[t]{0.45\textwidth}
\centering
\includegraphics[width=\textwidth]{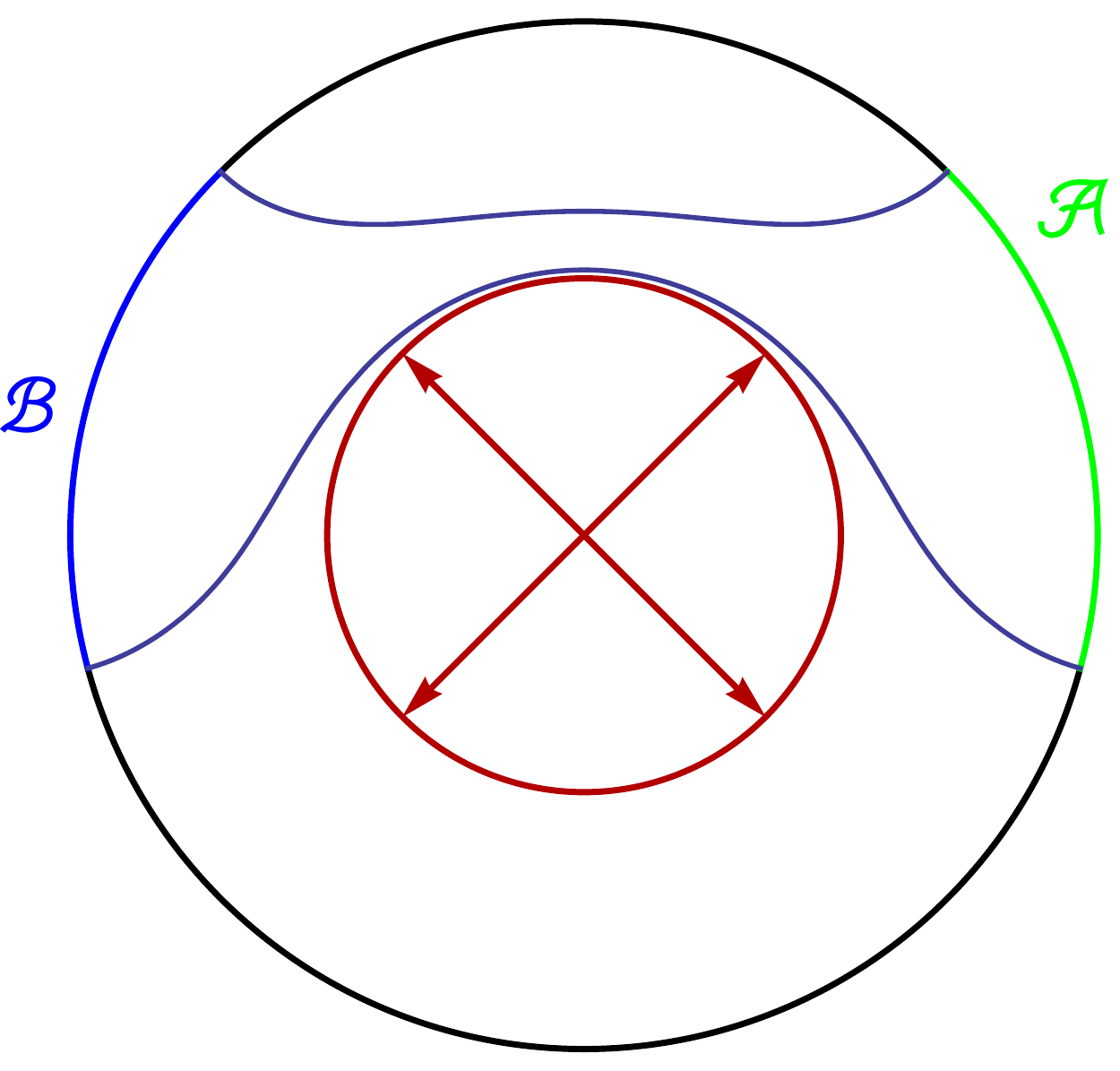}
\caption{Phase 2. Dominates for small $\Delta$ if $l+\Delta<\pi$.}
\end{subfigure}\\
\begin{subfigure}[t]{0.45\textwidth}
\centering
\includegraphics[width=\textwidth]{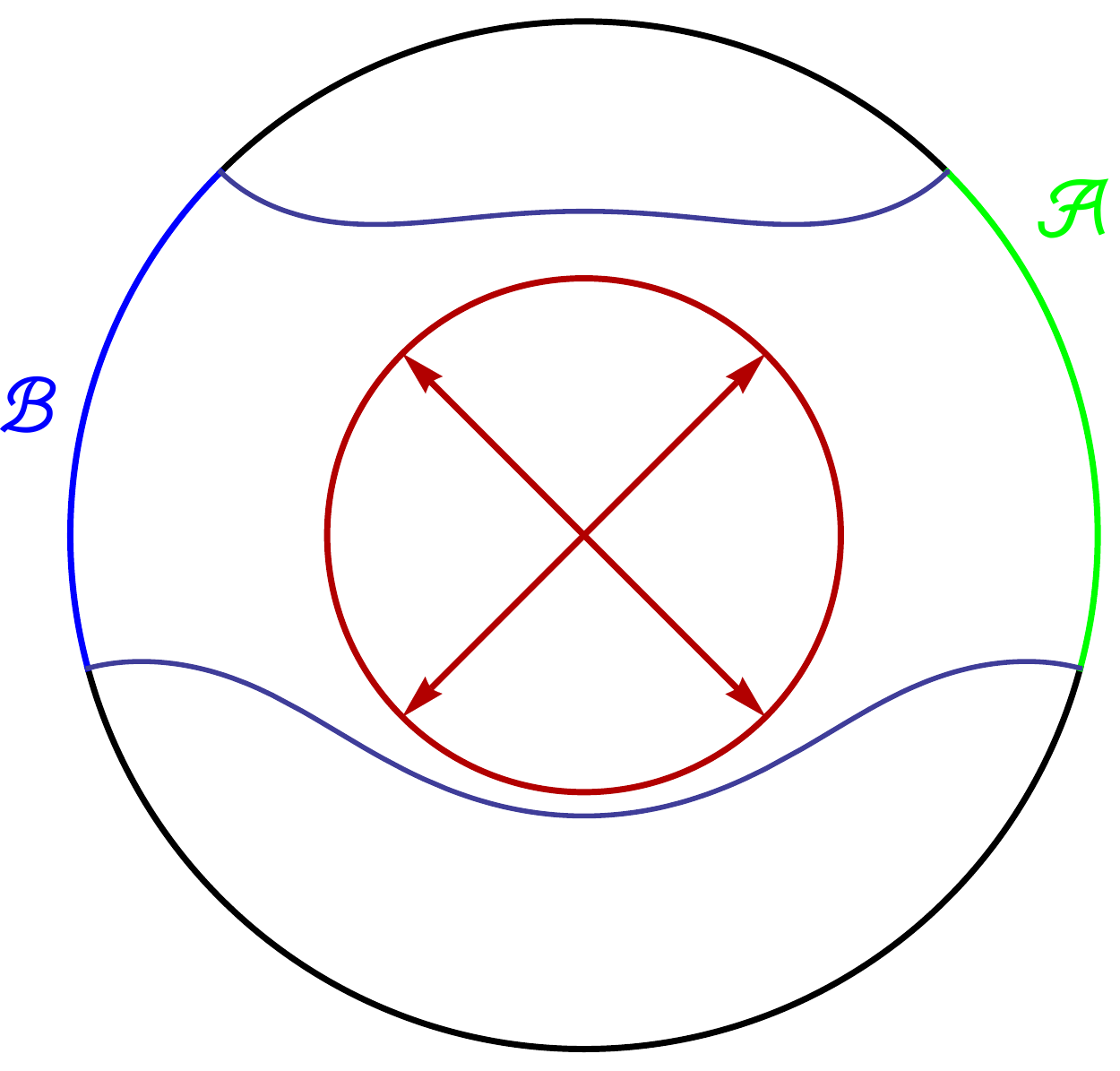}
\caption{Phase 3. Dominates for small $\Delta$ if $l+\Delta>\pi$.}
\end{subfigure}
\hspace{.07\textwidth}
\begin{subfigure}[t]{0.45\textwidth}
\centering
\includegraphics[width=\textwidth]{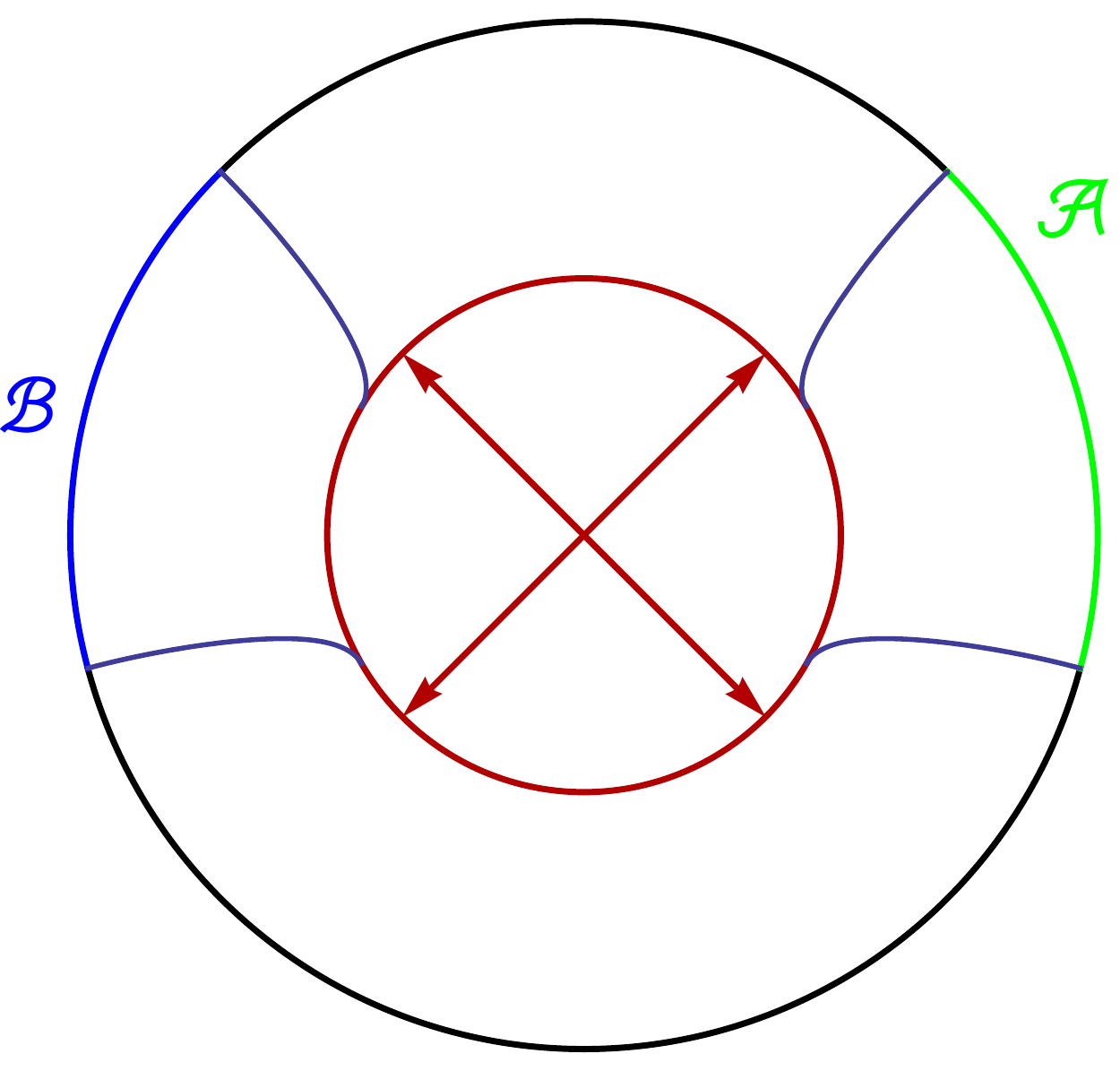}
\caption{Phase 4. May dominate in some intermediate regime.}
\end{subfigure}
\caption{The choices of geodesics to compute $S(\mathcal{A}\cup\mathcal{B})$ in the $\RR\PP^2$ geon. Options (a) and (b) would be admissible in BTZ, and (c) would additionally require inclusion of the event horizon. The geodesics in (d) pass through the crosscap, the arrows indicating that the antipodal points are identified.}\label{fig:RP2Geodesics}
\end{figure}

More interesting is the question of the mutual information $I(\mathcal{A}:\mathcal{B})=S(\mathcal{A})+S({\mathcal{B}})-S(\mathcal{A}\cup\mathcal{B})$ of two intervals $\mathcal{A}$ and $\mathcal{B}$. Consider the special case where the intervals have equal lengths $l<\pi$, with the centres of the intervals separated by $l<\Delta<\pi$. For the thermal state, this would have three different possible phases. The purity of the state changes one of these, much as in the case of a single interval, by removing the necessity of including the event horizon. Perhaps more interestingly, there is now an additional fourth phase, where the two geodesics pass through the centre. No closed geodesic is required to satisfy the homology constraint, so the argument that these are not shortest does not apply, and in fact for large enough $\rp$ it turns out that this phase can dominate. The geodesics relevant to the four phases\footnote{Interpreting phases as different OPE channels in twist operator correlation functions, one might object to a geodesic connecting the left end of one interval to the right end of another, since nonvanishing charges of the twist operators do not allow the vacuum block to appear in the OPE expansion. But this is circumvented here, since the operators should be brought together through an antiholomorphic transition map, reversing the direction of the twist for one operator.} are shown in \cref{fig:RP2Geodesics}.

\begin{figure}
\centering
\begin{subfigure}[t]{0.4\textwidth}
\centering
\includegraphics[width=\textwidth]{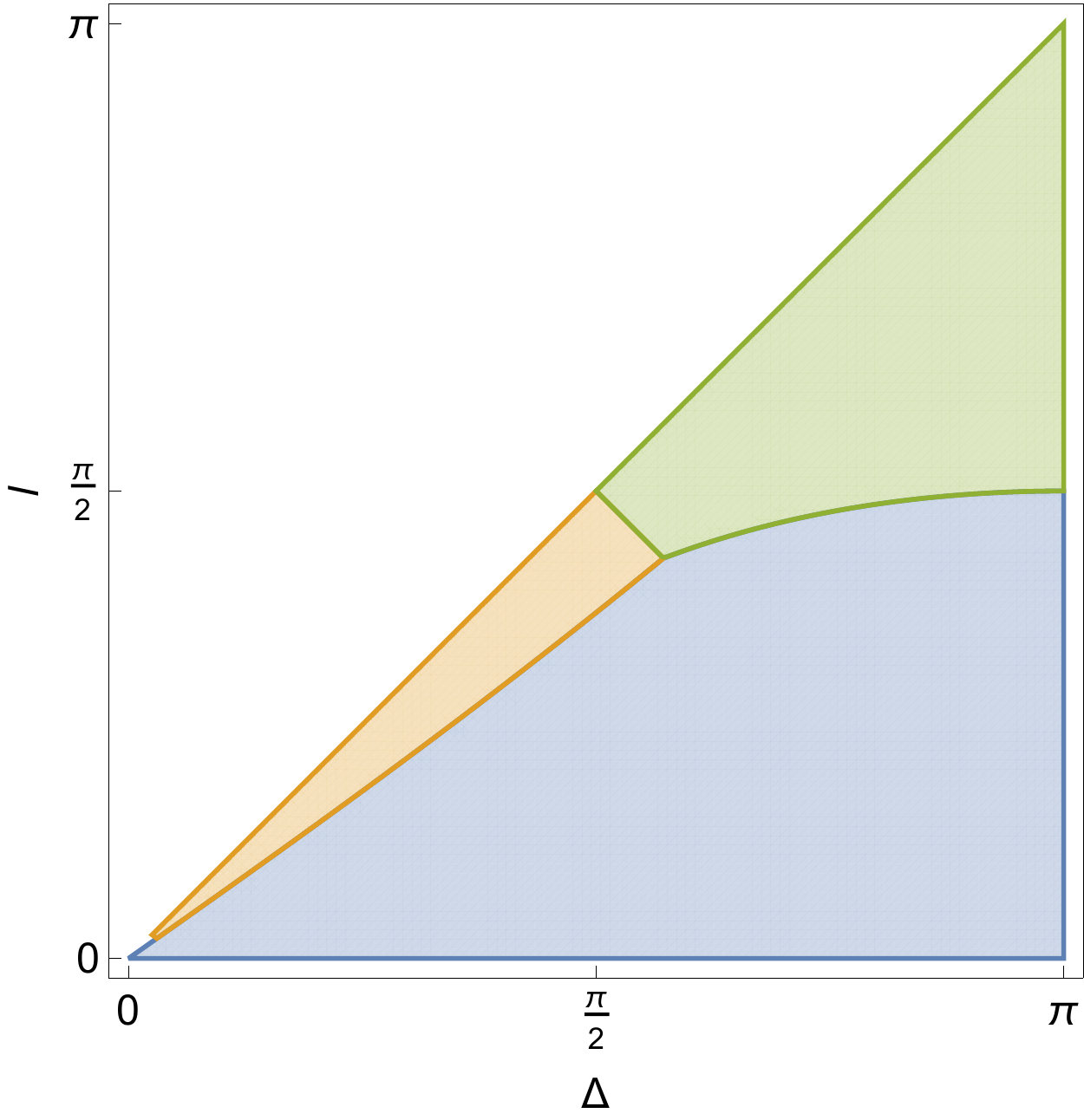}
\caption{$\rp=1$}
\end{subfigure}
\hspace{.07\textwidth}
\begin{subfigure}[t]{0.4\textwidth}
\centering
\includegraphics[width=\textwidth]{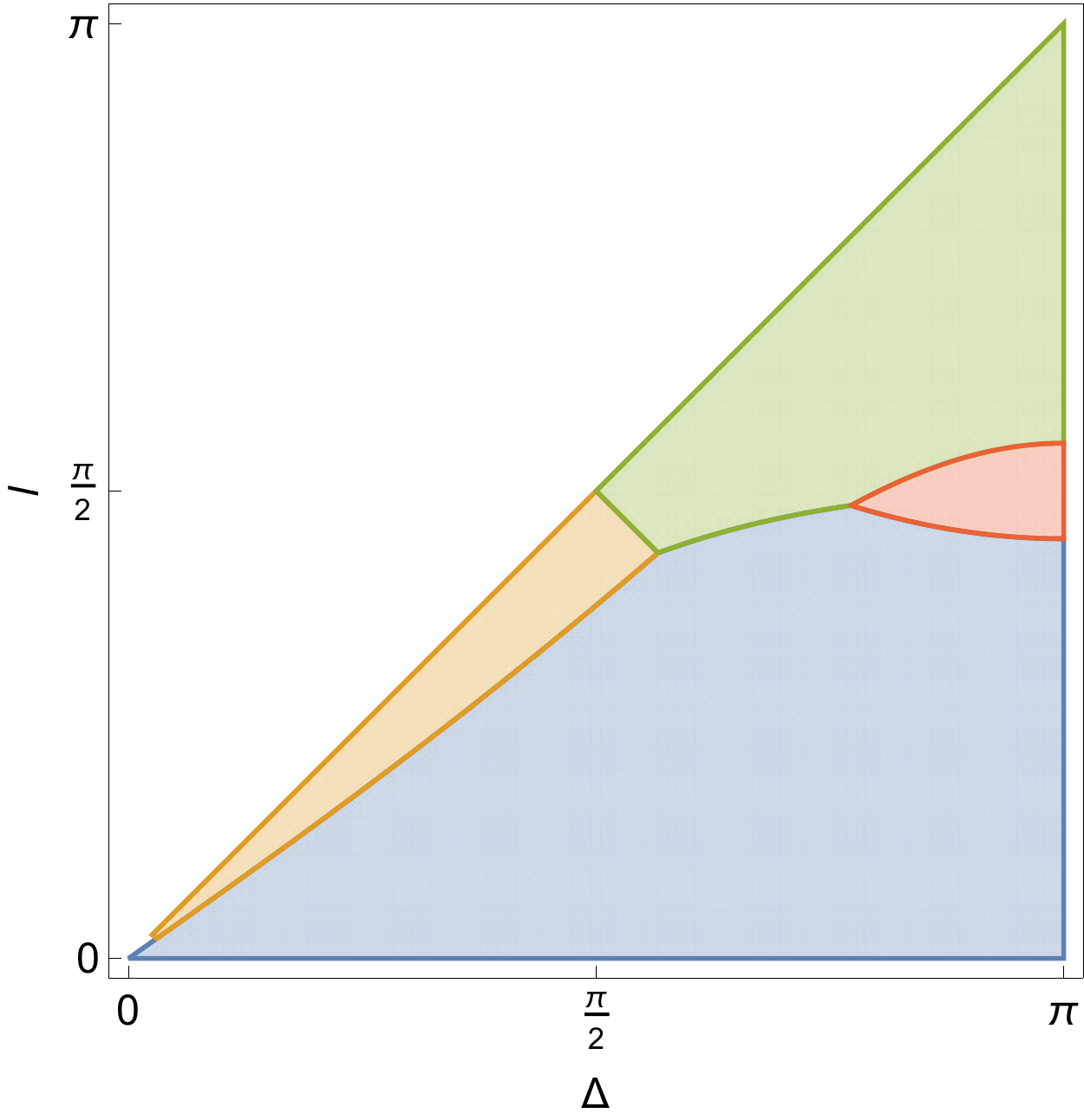}
\caption{$\rp=1.25$}
\end{subfigure}\\
\begin{subfigure}[t]{0.4\textwidth}
\centering
\begin{overpic}[width=\textwidth]{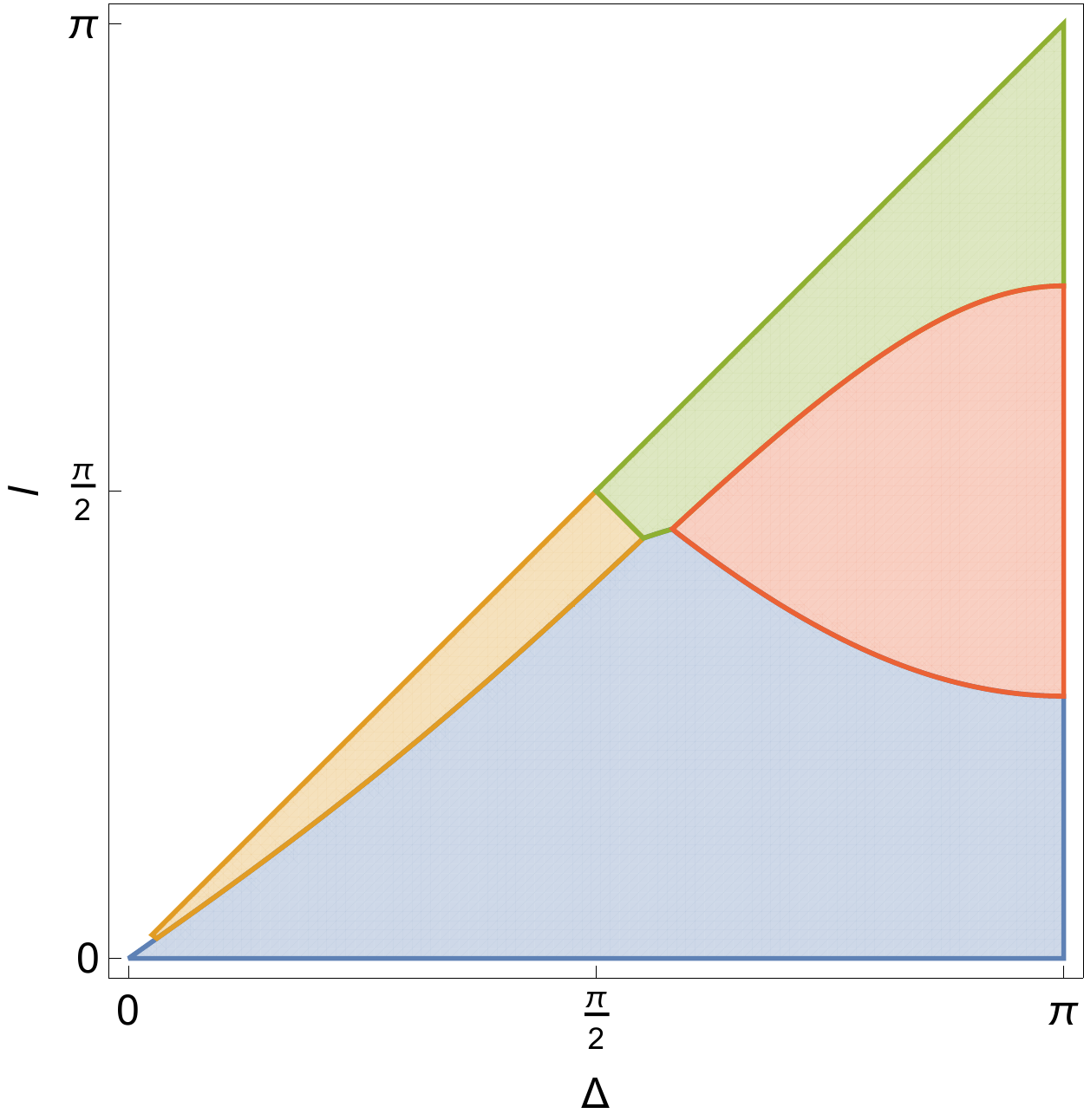}
\put(60,30){$1$}
\put(37,36){$2$}
\put(78,75){$3$}
\put(80,52){$4$}
\end{overpic}
\caption{$\rp=2$}
\end{subfigure}
\hspace{.07\textwidth}
\begin{subfigure}[t]{0.4\textwidth}
\centering
\includegraphics[width=\textwidth]{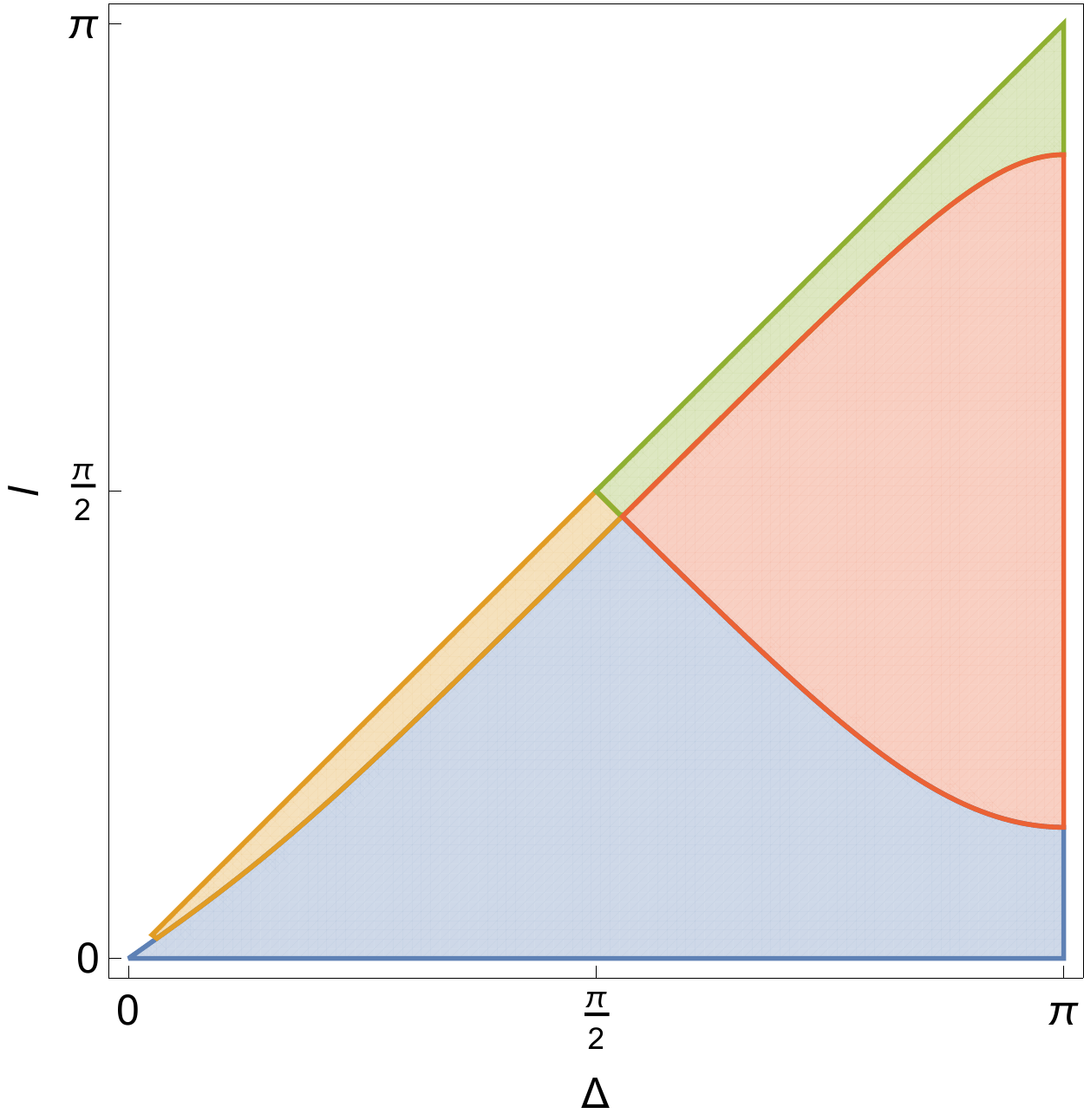}
\caption{$\rp=4$}
\end{subfigure}
\caption{Phases of mutual information of two intervals in the $\RR\PP^2$ geon, of length $l$, plotted vertically, with centres separated by $\Delta$, plotted horizontally,  for various values of $\rp$. The blue, yellow, green and red regions indicate where phases 1,2,3 and 4 respectively dominate, as labelled in subfigure (c).}\label{fig:RP2Phases}
\end{figure}

At $t=0$, the results for the mutual information in the different phases are
\begin{align*}
I_1 &= 0 \\
I_2 &= 2\log\left(\frac{\sinh ^2\left(\frac{l \rp}{2}\right)}{\sinh\left(\frac{\rp}{2}(\Delta-l)\right)\sinh\left(\frac{\rp}{2}(\Delta+l)\right)}\right) \\
I_3 &= 2\log\left(\frac{\sinh ^2\left(\frac{l \rp}{2}\right)}{\sinh\left(\frac{\rp}{2}(\Delta-l)\right)\sinh\left(\frac{\rp}{2}(2\pi-\Delta-l)\right)}\right) \\
I_4 &= 4\log\left(\frac{\sinh\left(\frac{l \rp}{2}\right)}{\cosh\left(\frac{\rp}{2}(\pi-\Delta)\right)}\right)
\end{align*}
with the mutual information equal to the largest of the four. The most interesting fourth phase is most likely to dominate when the separation $\Delta$ is at its largest possible value of $\pi$, and the lengths $l$ of the intervals are $\pi/2$. In this case we have $0=I_1=I_3>I_2$ and $I_4=4\log\sinh\left(\frac{\rp \pi}{4}\right)$, so the upshot is that $I_4$ can dominate when $\rp>\frac{4}{\pi}\sinh^{-1}1\approx 1.12$. The regions in parameter space where the four phases are dominant are shown for various values of $\rp$ in \cref{fig:RP2Phases}.

To understand the new phase, it helps to consider taking a large $\rp$ limit. For BTZ, this high temperature limit destroys correlations between all regions that are not parametrically close to the edge of parameter space, so that phase 1 dominates and the mutual information always vanishes. This is not the case in the $\RR\PP^2$ geon, for which there are correlations for sufficiently large or well separated regions with $l+\Delta>\pi$, when the fourth phase dominates:
\begin{equation}
I \sim \max\{0,2(l+\Delta-\pi)\rp\}
\end{equation}
In particular, for small regions the mutual information is nonzero only when $\Delta$ is very close to $\pi$. This suggests that a region is most strongly entangled with regions furthest from it. It would be interesting to understand how this arises from the state of the field theory.

A final exercise is to consider what happens to the mutual information under time evolution. This does not alter the first three phases, but the fourth phase changes to
\begin{equation}
I_4= 2\log\left(\frac{\sinh^2\left(\frac{l \rp}{2}\right)}{\cosh\left(\frac{\rp}{2}(\pi-\Delta+2t)\right)\cosh\left(\frac{\rp}{2}(\pi-\Delta-2t)\right)}\right).
\end{equation}
Despite the fact that the external region is static, the entanglement entropy is sensitive to the non-staticity behind the horizon. The results are indicative of a very special state at $t=0$, which thermalizes after some order 1 time.
Focussing on the intervals for which the fourth phase is most dominant, namely $\Delta=\pi$ and $l=\pi/2$, we see that the mutual information is positive when $\sinh\left(\frac{\pi\rp}{4}\right)>\cosh(\rp t)$. The mutual information declines and saturates to zero at some time, which is in fact always bounded by $\pi/4$, for all $\rp$.

In the large $\rp$ limit, $I_4$ has a cross over at $|t|=(\pi-\Delta)/2$, when the geodesic moves from hugging the horizon to staying close to the singularity. Suggestively, this is exactly the time at which all right-movers from $\mathcal{A}$ have had a chance to meet some left-movers from \emph{directly opposite} $\mathcal{B}$, the region $\mathcal{B}$ is most strongly entangled with according to the previous analysis.
It begins to decrease linearly, as $2(l-2|t|)\rp$ until saturating to either zero mutual information at time $l/2$, or to phase 3 at time $(\pi-l)/2$ for sufficiently large regions $l>\pi/2$. It would be interesting to attempt to understand whether this can be interpreted from a quasi-particle picture of the dynamics of entanglement, at least in this limit, and what this implies for the way the entanglement is distributed in the CFT state.

\subsection{Three boundary wormhole}\label{sec:wormhole3}

Before beginning with the cases of quotient groups generated by two elements, we describe a useful way to understand the $SL(2,\RR)$ group structure, by identifying elements with the tangent space at the identity via the exponential map. For simplicity, we restrict here to the case of nonrotating wormholes, for which the quotient group lies in the diagonal $PSL(2,\RR)$. The generalisation to the rotating case is relatively straightforward.

Write the generators as exponentials of $\mathfrak{sl}(2,\RR)$ Lie algebra elements, parameterised by
\begin{equation}
\xi=\frac12
\begin{pmatrix}
z & x-t \\ x+t & -z
\end{pmatrix}.
\end{equation}
This three dimensional Lie algebra has a natural Lorentzian inner product from the Killing form, calculable from the determinant, and $PSL(2,\RR)$ isometries act by conjugation, giving the three dimensional Lorentz transformations.
In the picture of $SL(2,\RR)$ as $AdS_3$, this Lorentzian structure is inherited from the tangent space of the origin.

 The hyperbolic isometries are given by exponentials of spacelike elements. Thinking of the isometries like this as vectors in $\RR^{2,1}$, with the coordinates $(t,x,z)$, one generator of $\Gamma$ may  be boosted and rotated to lie along the $x$-axis. This leaves a single residual symmetry of boosting in the $z$ direction. Depending on the second generator, there are then two distinct cases depending on whether its $z$ component or $t$ component can be boosted to zero\footnote{The third possibility, that the residual vector is null, means that $\Gamma$ contains parabolic, if not elliptic, elements. We will not consider it here.}.

In the case when the $x$ component can be set to zero, the spacetime that results is a wormhole with three asymptotic boundaries, all connected through a non-traversable bridge. We begin in this section with this class. In the opposing case when the $t$ component may be set to zero, the result is a wormhole with a single exterior region and a torus behind the event horizon, which we move to in the next section (\cref{sec:Torus}).

There are three moduli to specify the spacetime, which can be picked in the Lorentzian Lie algebra language as the two lengths and the boost angle between them. A convenient choice of the generators is $g_1=\exp\xi_1$ and $g_2=\exp\xi_2$, where the $\mathfrak{sl}(2,\RR)$ Lie algebra elements are given by
\begin{equation}
\xi_1=
\frac{\ell_1}{2}
\begin{pmatrix}
0 & 1 \\ 1 & 0
\end{pmatrix},
\quad
\xi_2=
\frac{\ell_2}{2}
\begin{pmatrix}
0 & e^\alpha \\ e^{-\alpha} & 0
\end{pmatrix}
\end{equation}
so that
\begin{equation}
g_1=
\begin{pmatrix}
\cosh\left(\frac{\ell_1}{2}\right) & \sinh\left(\frac{\ell_1}{2}\right) \\
 \sinh\left(\frac{\ell_1}{2}\right) & \cosh\left(\frac{\ell_1}{2}\right)
\end{pmatrix},
\quad
g_2=
\begin{pmatrix}
\cosh\left(\frac{\ell_2}{2}\right) & e^\alpha \sinh\left(\frac{\ell_2}{2}\right) \\
 e^{-\alpha}\sinh\left(\frac{\ell_2}{2}\right) & \cosh\left(\frac{\ell_2}{2}\right)
\end{pmatrix}.
\end{equation}

The $\xi$s generate translations in the first and second asymptotic regions, and the geodesics connecting the fixed points of $g_1$ and $g_2$ lie along the event horizons of these regions. The corresponding element of $\Gamma$ for the third region in this parametrization is $g_3=-g_1 g_2^{-1}$. For the spacetime to be free of conical singularities, this generator must also be hyperbolic, which is equivalent to the condition that the boundaries of the fundamental region on the $t=0$ slice do not meet. This is achieved for sufficiently large $\alpha$, such that $e^\alpha >\coth\left(\frac{\length_1}{4}\right)\coth\left(\frac{\length_2}{4}\right)$, and when this is satisfied the length of the third event horizon can be found from \eqref{eq:closedLength}:
\begin{align}
e^\alpha&=\csch\left(\frac{\ell_1}{2}\right)\csch\left(\frac{\ell_2}{2}\right) \Bigg[ \cosh \left(\frac{\ell_3}{2}\right)+\cosh \left(\frac{\ell_1}{2}\right) \cosh \left(\frac{\ell_2}{2}\right) +  \\ \nonumber
& 
  \left(2 \cosh \left(\frac{\ell_1}{2}\right) \cosh \left(\frac{\ell_2}{2}\right) \cosh \left(\frac{\ell_3}{2}\right)+\frac{\cosh (\ell_1)+\cosh
   (\ell_2)+\cosh (\ell_3)+1}{2}\right)^{1/2} \Bigg]
\end{align}

\begin{figure}
\centering
\includegraphics[width=0.4\textwidth]{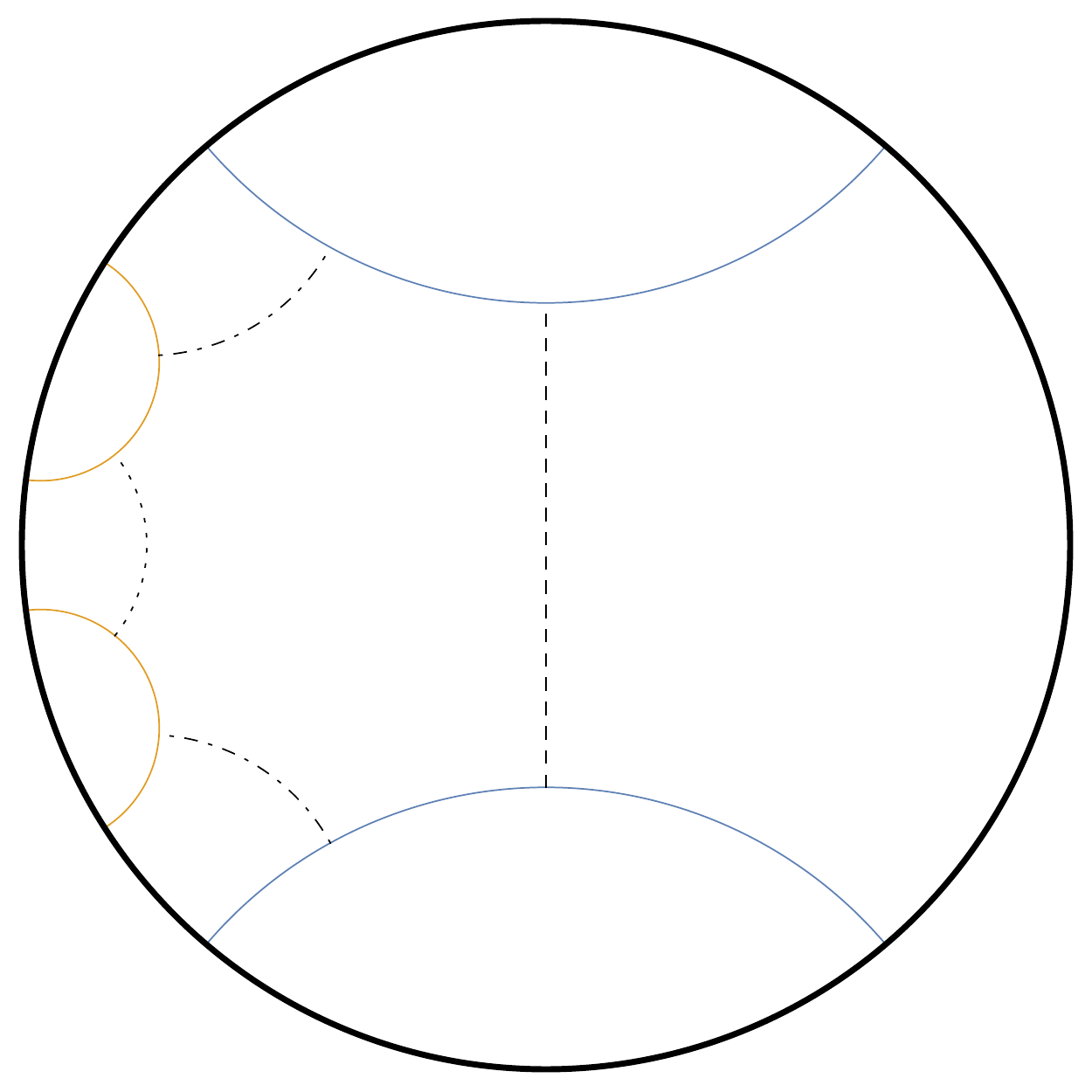}
\caption{The $t=0$ slice of $AdS_3$, showing a fundamental region and event horizons for the three boundary wormhole. The blue curves, identified by $g_1$, and the orange, identified by $g_2$, mark the edge of the fundamental domain. The dashed, dotted, and dot-dashed lines mark the event horizons of the three exterior BTZ regions.}\label{fig:3wormholePoincare}
\end{figure}

A symmetric choice of fundamental region on the $t=0$ slice is shown in \cref{fig:3wormholePoincare}, along with the geodesics that lie on the event horizons for each exterior region.

In the CFT, this quotient is (in some region of moduli space where this phase dominates) dual to a pure entangled state on three noninteracting circles, prepared by a path integral over a Riemann surface with three boundaries and no handles. Joining this to its reflection on the boundaries, to form the Schottky double as described in \cref{sec:EuclidLorentz}, one obtains a closed Riemann surface of genus two.

Just as in the case of BTZ, this genus two surface is a boundary to many different on-shell Euclidean bulk spaces. There are five choices of handlebody that respect the time-reversal symmetry, and hence allow for translation back into Lorentzian language, and can be distinguished by connectedness of the Lorentzian section, exactly as for the thermal state. The physics will be described by the dominant phase with the smallest Euclidean action, which will depend on the 3 real moduli of the Riemann surface. One phase gives three disconnected copies of $AdS$, so there should be no correlations between the copies of the CFT in the classical limit, and 3 phases have one boundary disconnected and the others joined by a BTZ black hole. The final fifth phase is our wormhole geometry.  It is an open question to determine the moduli for which each phase dominates, so excepting for certain symmetric situations and limits we do not know whether the wormhole is relevant. See \cite{Balasubramanian:2014hda} for a more detailed discussion. From here on, we will assume that we are always in the fully connected wormhole phase.

Here, we will look only at entanglement entropy for a single interval in one of the asymptotic regions. There are at least two phases that must inevitably dominate for some regions of the parameter space, analogous to the two phases for small and large intervals in the thermal state. In these cases, the geodesics remain outside the horizon in the region isometric to BTZ, and differ only in deciding which way to pass round the horizon. The closed geodesics required in the case of the `large region' regime may live either on the horizon of the region in question, or on both the other two horizons if the sum of these lengths is smaller\footnote{There exist many more complicated closed geodesics that suffice to satisfy the homology constraint, but a thorough search as described later has shown them to always be longer than the horizons. This is unsurprising, though we do not know of any proof that this must always hold.}. The more interesting possibility is whether there may be intermediate phases that ever dominate, with geodesics meeting the boundary and yet passing through the interior region. It turns out that (at least ignoring the question of which bulk geometry phase dominates) this possibility is realised.

We begin again by choosing representative points in each boundary region. Due to the symmetric way in which we have picked our generators, this is very easy to do in the first and second asymptotic regions, though rather harder in the third. We will not reproduce the results for the third region here, though there will be very similar difficulties encountered in the next section, which will serve to illustrate how to proceed if required.

From \cref{fig:3wormholePoincare}, it is clear that a simple choice of reference points in the first an second regions will be $\vec{u}=\vec{v}\propto(1,0)^t$ and $\vec{u}=\vec{v}\propto(0,1)^t$ respectively, lying at the far right and left of the disc. Normalising these, and translating with the appropriate Killing vectors, just as for the BTZ example, we obtain
\begin{equation}
\vec{u}_1=\vec{v}_1=\sqrt{\frac{4 \pi }{\ell_1}} 
\begin{pmatrix}\sinh\left(\frac{\ell_1\phi}{4\pi}\right)\\
\cosh\left(\frac{\ell_1\phi}{4\pi}\right)\end{pmatrix},\quad
\vec{u}_2=\vec{v}_2=\sqrt{\frac{4 \pi }{\ell_2}} 
\begin{pmatrix}\cosh\left(\frac{\ell_2\phi}{4\pi}\right)\\
e^{-\alpha}\sinh\left(\frac{\ell_2\phi}{4\pi}\right)\end{pmatrix}
\end{equation}
restricting to time $t=0$ (the generalisation to include time dependence is exactly as for BTZ). We will use only the result for the first region here, to compute the entanglement entropy of an interval of length $\Delta\phi$, centred at angle $\phi_0$, so between $\phi=\phi_0 \pm \frac{\Delta\phi}{2}$. By symmetry, we need only consider $\phi\in [0,\pi]$, and $\Delta\phi\in (0,2\pi)$. The extremes of $\phi_0=0,\pi$ are when the interval is closest to the second and third asymptotic regions respectively.

The length results are quick to obtain from \cref{eq:openLength} as before, though the exact forms are unrevealing so we will not reproduce them here. One advantage of having a systematic way of finding a relatively simple analytic answer is that it allows for computing a very large number of lengths for many different homotopy classes of geodesic. In the present case, we examined lengths of all geodesics with homotopy classes formed from words in the generators with up to eight letters, for a random choice of over one million sets of parameters, checking which classes dominated in any case, using \texttt{Mathematica} \cite{mathematica10}. This was performed by first creating a list of all the elements of the group formed from some finite number of generators, and then calculating the appropriate matrix traces analytically. Each of these analytic expressions was turned into a C-compiled function of the parameters $(\ell_1,\ell_2,\ell_3,\Delta,\phi_0)$ for efficiency, and then evaluated on a randomly chosen set of these parameters repeatedly, checking the dominant geodesic each time. This gave a list of all realised possibilities in a few seconds, showing that there are only four phases of interest. An entirely similar calculation for closed geodesics (again checking conjugacy classes from words of up to eight letters in the generators) found that only the three event horizons are relevant.

\begin{figure}
\centering
\begin{subfigure}[t]{0.4\textwidth}
\centering
\includegraphics[width=\textwidth]{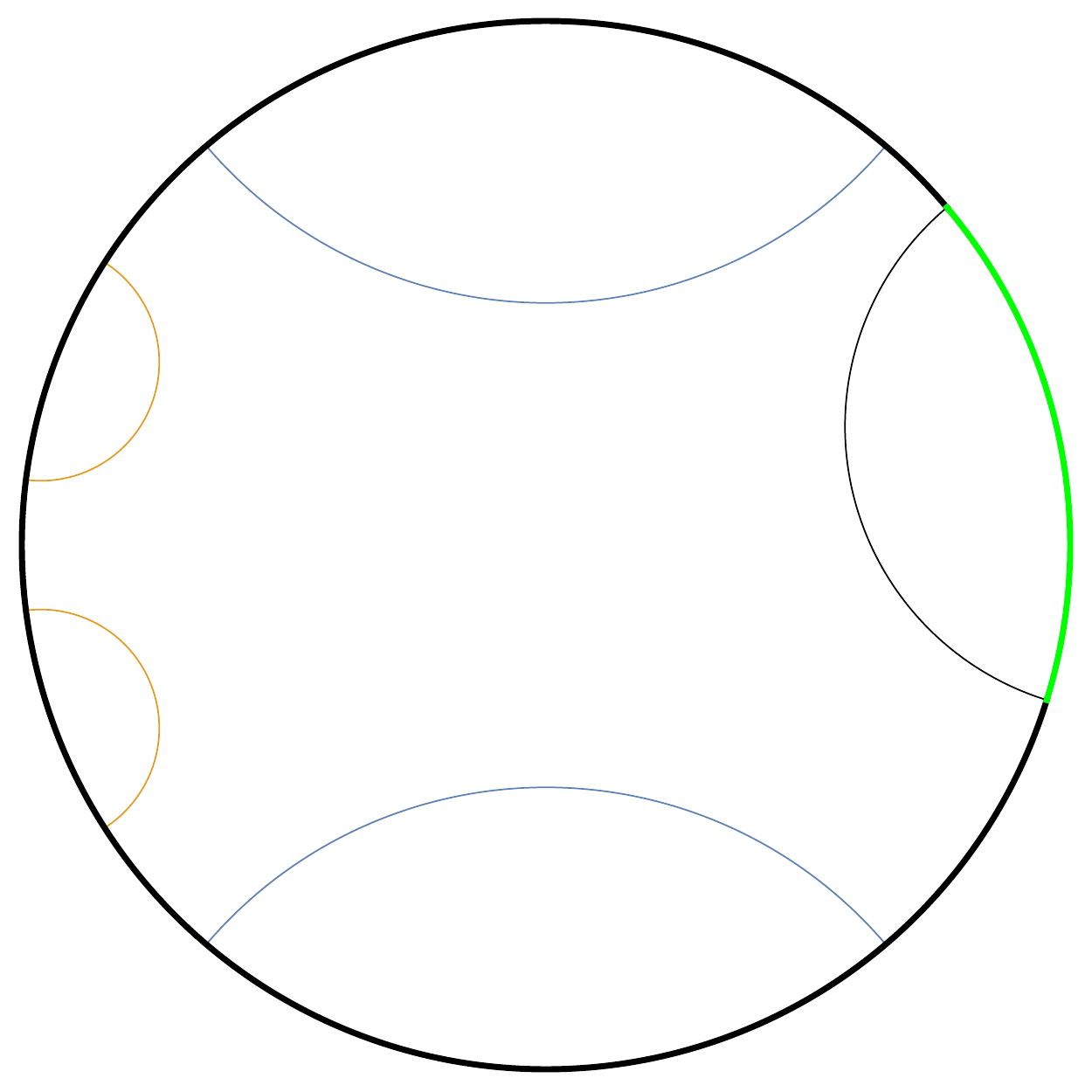}
\caption{Phase 1: $1$}
\end{subfigure}
\hspace{.07\textwidth}
\begin{subfigure}[t]{0.4\textwidth}
\centering
\includegraphics[width=\textwidth]{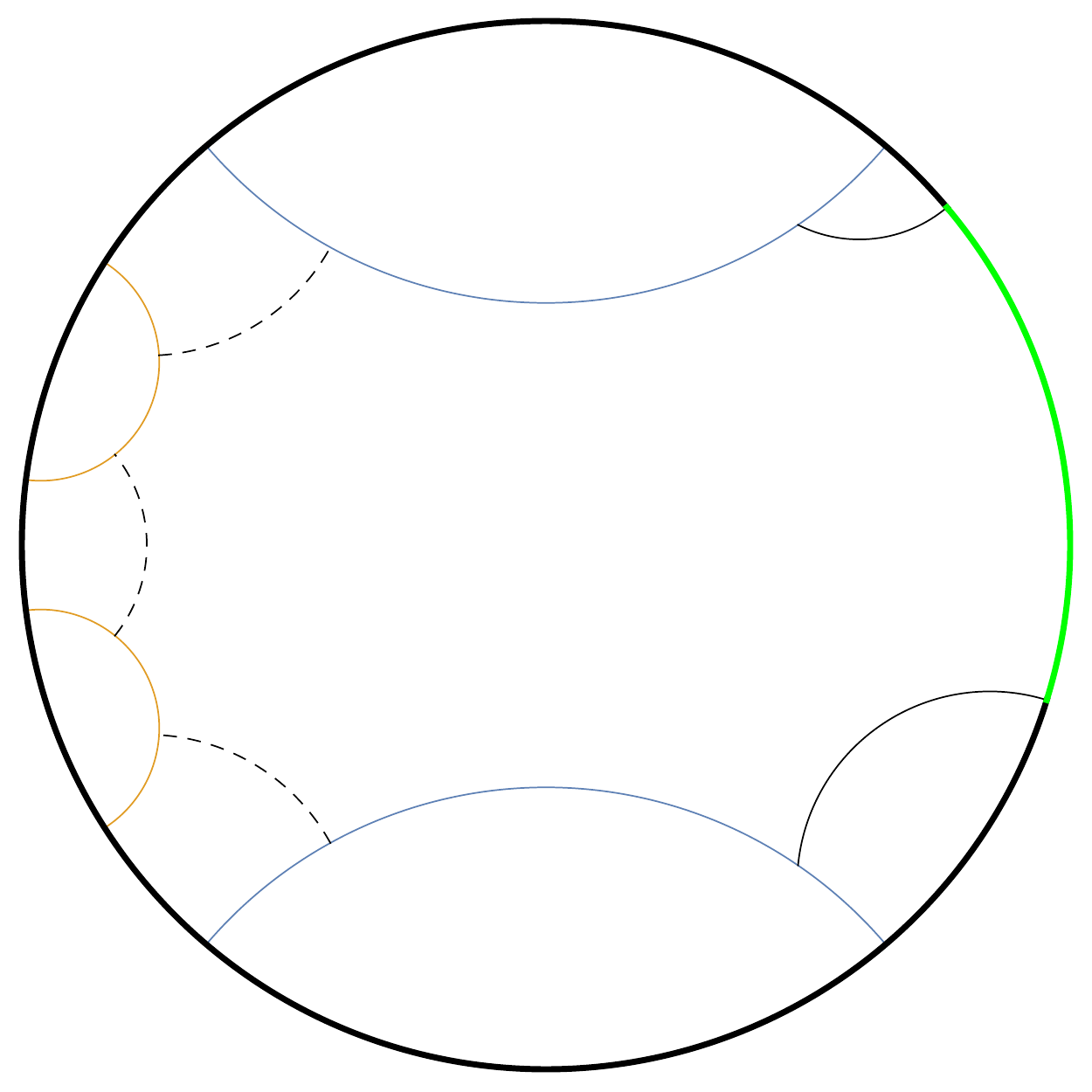}
\caption{Phase 2: $g_1^{-1}$}
\end{subfigure}\\
\begin{subfigure}[t]{0.4\textwidth}
\centering
\includegraphics[width=\textwidth]{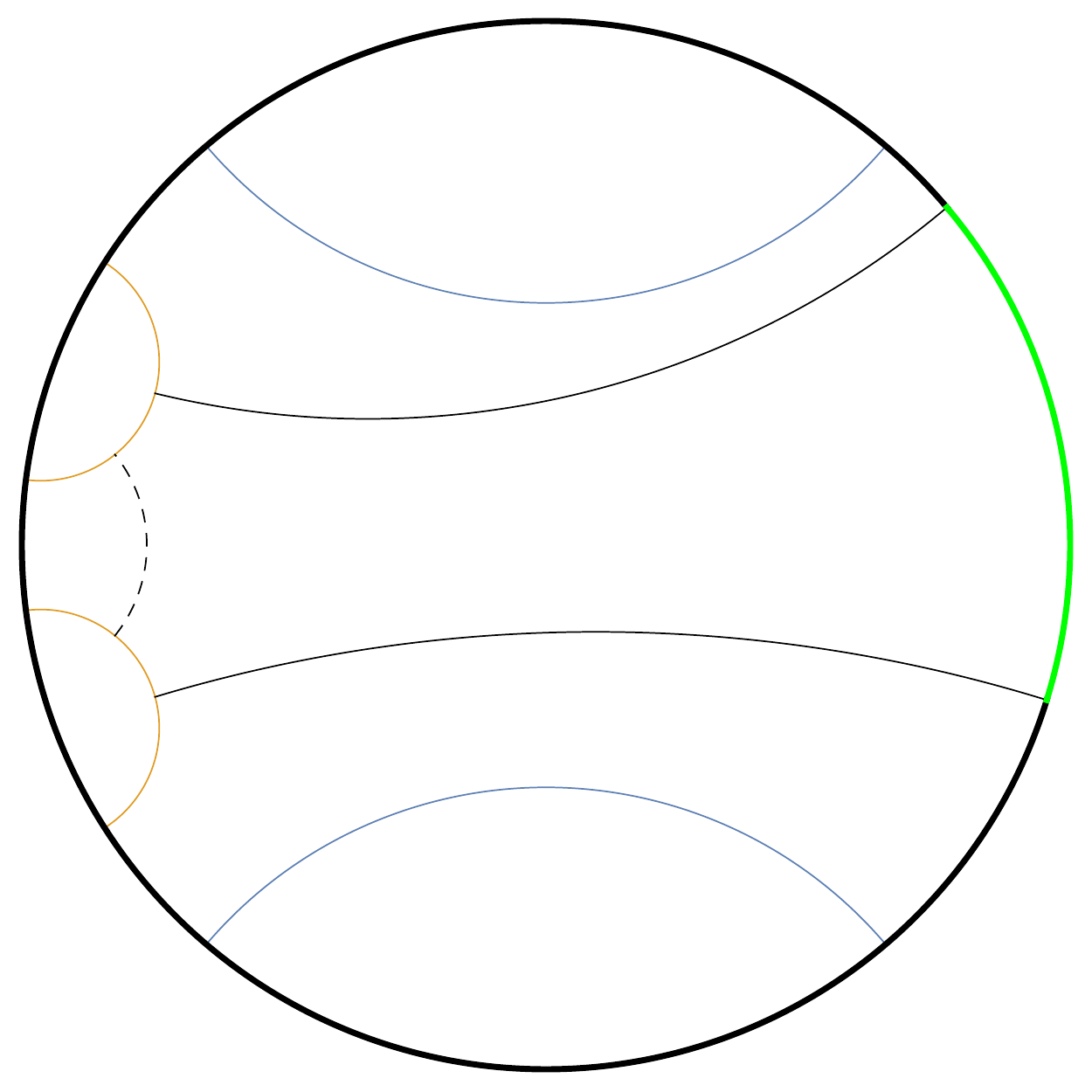}
\caption{Phase 3: $g_2^{-1}$}
\end{subfigure}
\hspace{.07\textwidth}
\begin{subfigure}[t]{0.4\textwidth}
\centering
\includegraphics[width=\textwidth]{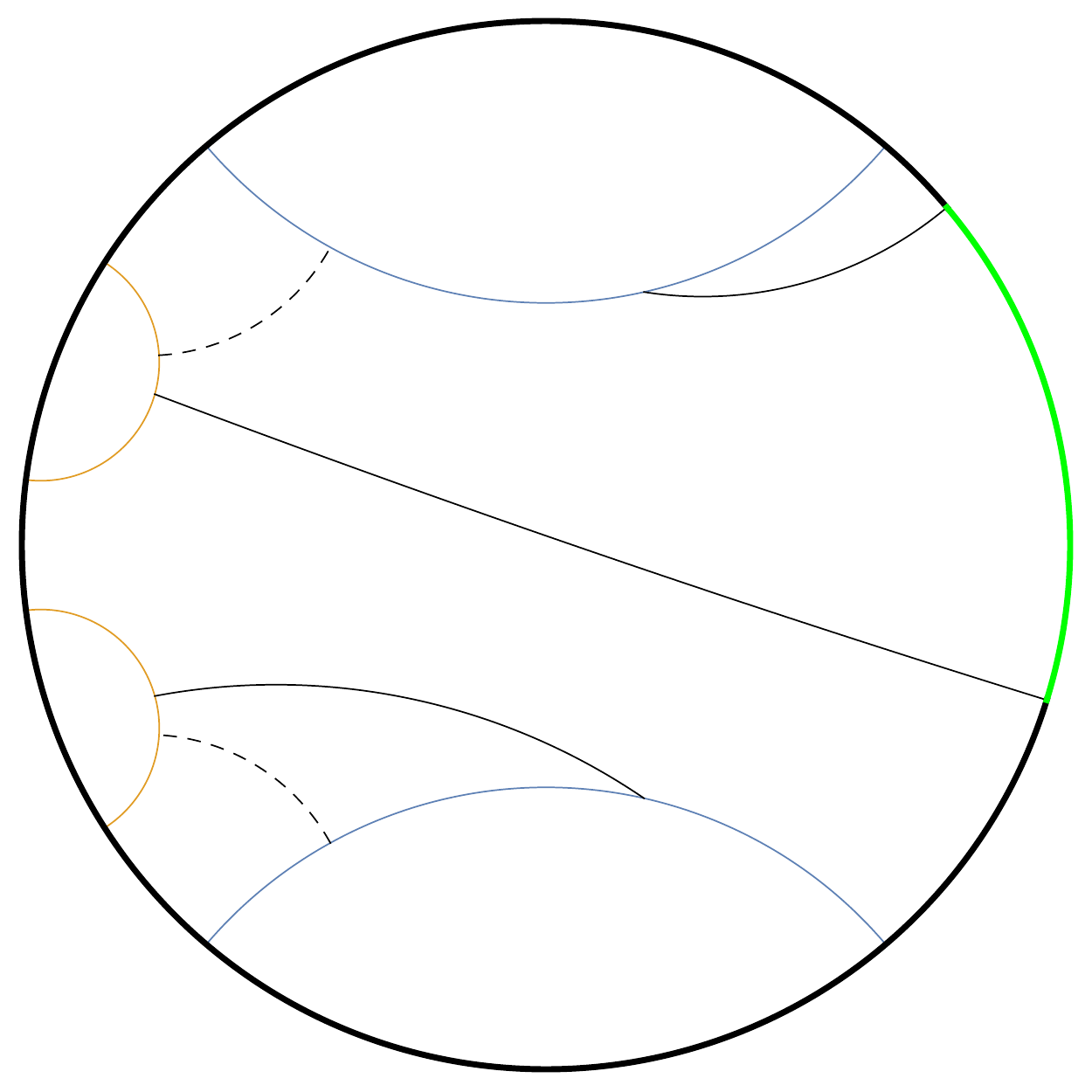}
\caption{Phase 4: $g_2g_1^{-1}$}
\end{subfigure}
\caption{The geodesics giving the four possible phases of entanglement entropy of a single interval, in green, along with the event horizons added to satisfy the homology constraint, marked by dashed lines.}\label{fig:3wormholePhases}
\end{figure}

The four phases, shown in \cref{fig:3wormholePhases}, correspond to the homotopy classes of the identity, $g_1^{-1}, g_2^{-1}$, and $g_3^{-1}=g_2 g_1^{-1}$, with the last three supplemented by the event horizons of exteriors 1, 2 and 3 (or a sum of the other two if shorter) respectively.

Each of the last three phases corresponds to saturation of some Araki-Lieb inequality \cite{Araki:1970ba}
\begin{equation}
|S(\mathcal{A})-S(\mathcal{B})|\leq S(\mathcal{AB}),\text{ saturation when} S(\mathcal{A})=S(\mathcal{B}) + S(\mathcal{AB})
\end{equation}
where $\mathcal{A}$ is the interval in question, and $\mathcal{B}$ is either the complement of $\mathcal{A}$ in boundary 1, as happens in the thermal state \cite{Hubeny:2013gta}, or the entirety of boundary 2 or 3. This has a very natural interpretation \cite{Xi2011} in terms of the state on $\mathcal{AB}$, that the Hilbert space of $\mathcal{A}$ can be split into two parts $\mathcal{A}_1$ and $\mathcal{A}_2$ in such a way that the state on $\mathcal{AB}$ factorizes as a mixed state on $\mathcal{A}_1$ times a pure state on $\mathcal{A}_1\mathcal{B}$. This means in this case that $\mathcal{B}$ is only entangled with some subset of degrees of freedom in $\mathcal{A}$, and the remaining degrees of freedom in $\mathcal{A}$ are entangled only with the remainder of the system.

\begin{figure}
\centering
\includegraphics[width=0.5\textwidth]{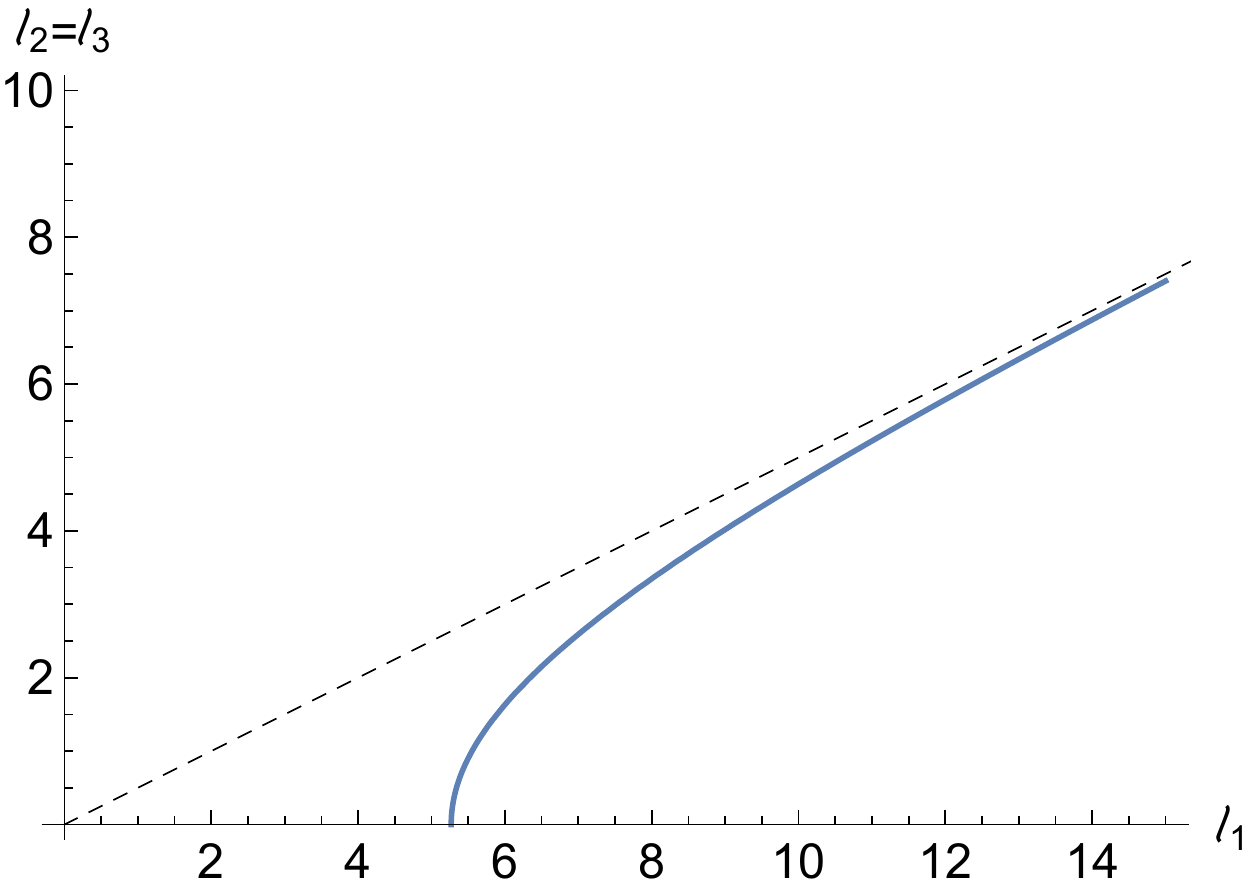}
\caption{Phase diagram for single-interval entanglement as a function of the moduli of the spacetime, in the case when two of the horizon lengths are equal. Below and to the right of the solid line, one of phases 3 or 4 dominates in some region of moduli space, and $S(\mathcal{A})$ has nontrivial dependence on space and time. This is always below the dashed line $\ell_1=\ell_2+\ell_3$, where there is a phase transition associated to the closed geodesics.}\label{fig:3wormholeLengthPhases}
\end{figure}

We find that the most interesting last two phases, where the geodesic passes behind the event horizon, may dominate as long as $\ell_1$ is larger than some order one value, and also somewhat larger than $\ell_2$ and $\ell_3$. We will focus here on the symmetric case when $\ell_2=\ell_3$, for which a phase diagram is plotted in \cref{fig:3wormholeLengthPhases}.
The result here is that for $\ell_1$ larger than some critical value, phases 2 and 3 dominate for some interval. For $\ell_{2,3}$ very small, the critical value of $\ell_3$ is some order one number, which then increases with $\ell_{2,3}$, approaching $2\ell_{2,3}$ for large horizons but never exceeding it. This means that when the entropy of region 1 is computed from its own horizon radius, rather than the sum of the other two, all single-interval entanglement entropies are in exact agreement with the thermal state. When the other two horizons are small, on the other hand, the two systems are insufficiently entropic to purify a thermal state in the first region. The entanglement entropy of a sufficiently large region notices that the state is not thermal, and may saturate its entanglement with one or other of the boundaries.

\begin{figure}
\centering
\begin{subfigure}[t]{0.4\textwidth}
\centering
\includegraphics[width=\textwidth]{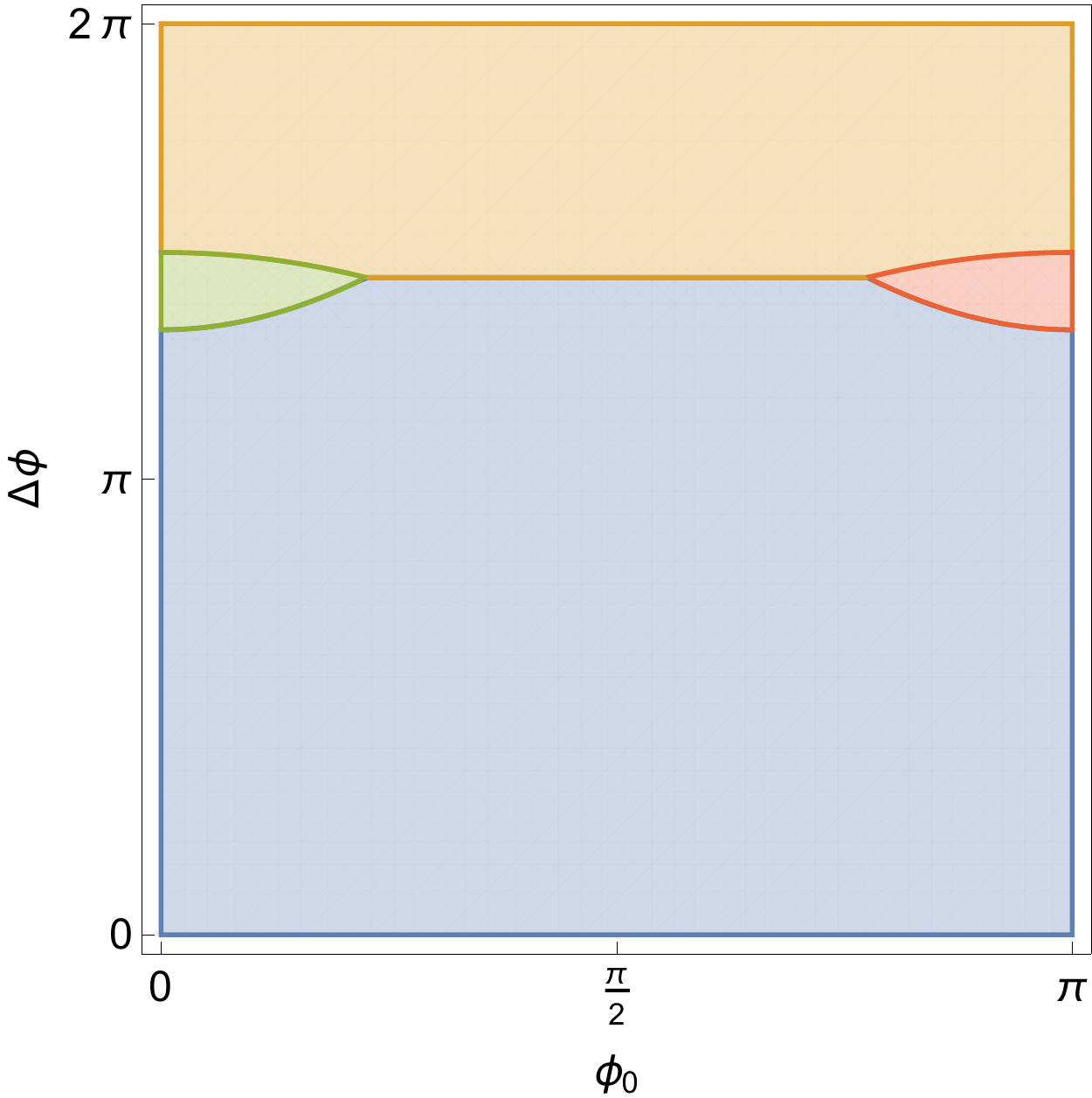}
\caption{$\ell_1=2\pi,\ell_2=\ell_3=\pi/2$}
\end{subfigure}
\hspace{.07\textwidth}
\begin{subfigure}[t]{0.4\textwidth}
\centering
\includegraphics[width=\textwidth]{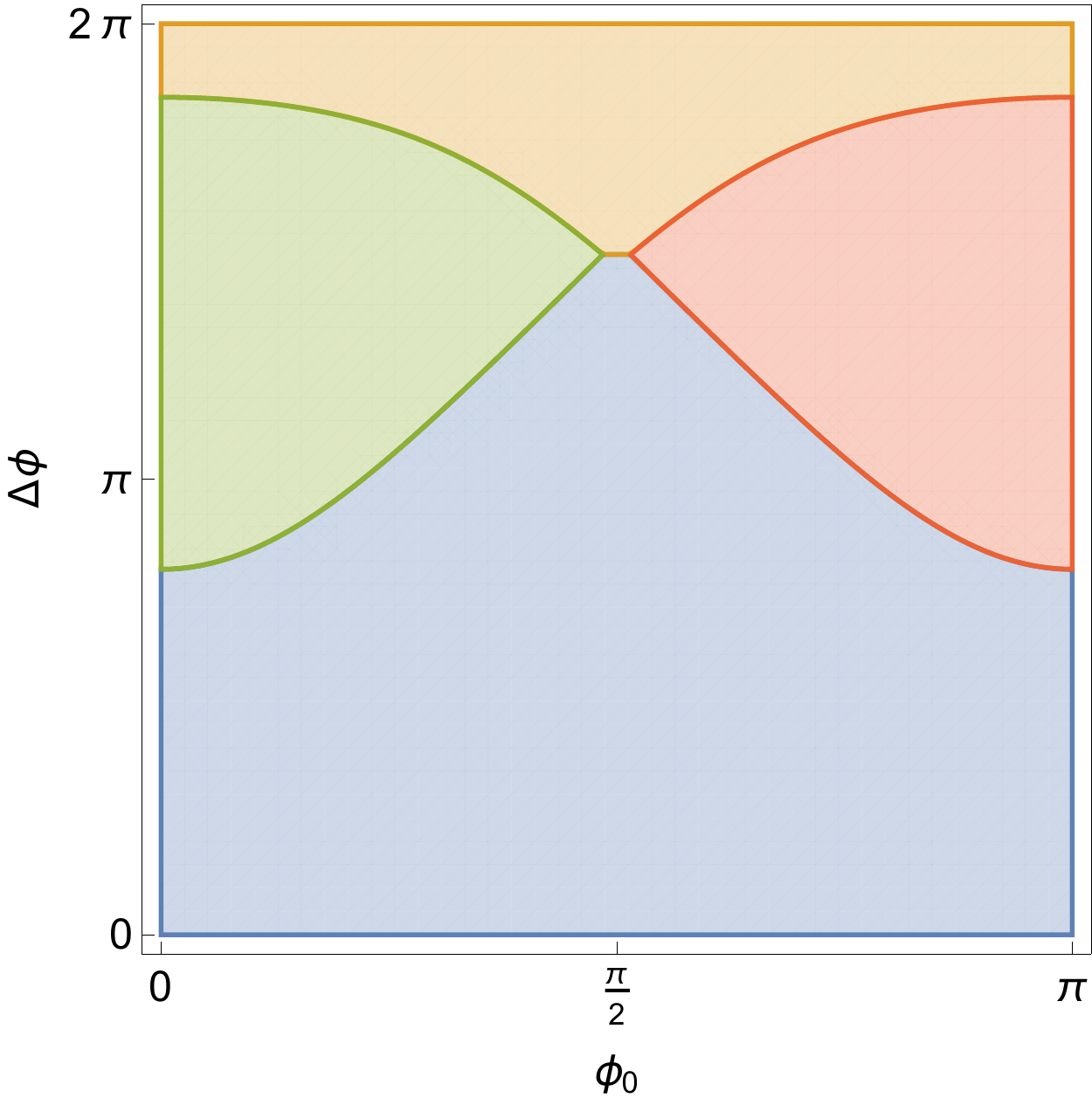}
\caption{$\ell_1=4\pi,\ell_2=\ell_3=\pi$}
\end{subfigure}\\
\begin{subfigure}[t]{0.4\textwidth}
\centering
\begin{overpic}[width=\textwidth]{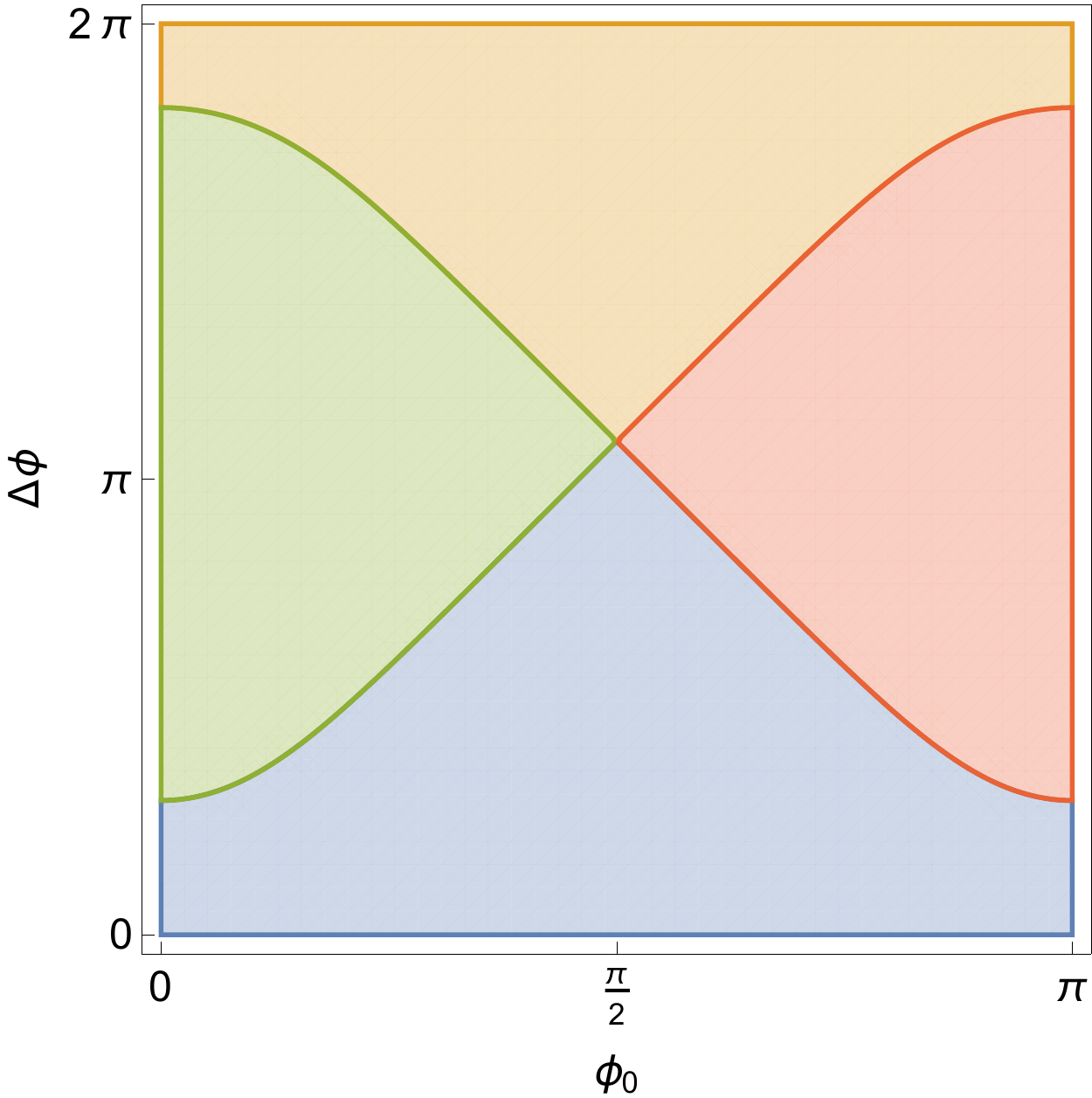}
\put(54,30){$1$}
\put(54,80){$2$}
\put(30,57){$3$}
\put(80,57){$4$}
\end{overpic}
\caption{$\ell_1=6\pi,\ell_2=\ell_3=\pi/4$}
\end{subfigure}
\hspace{.07\textwidth}
\begin{subfigure}[t]{0.4\textwidth}
\centering
\includegraphics[width=\textwidth]{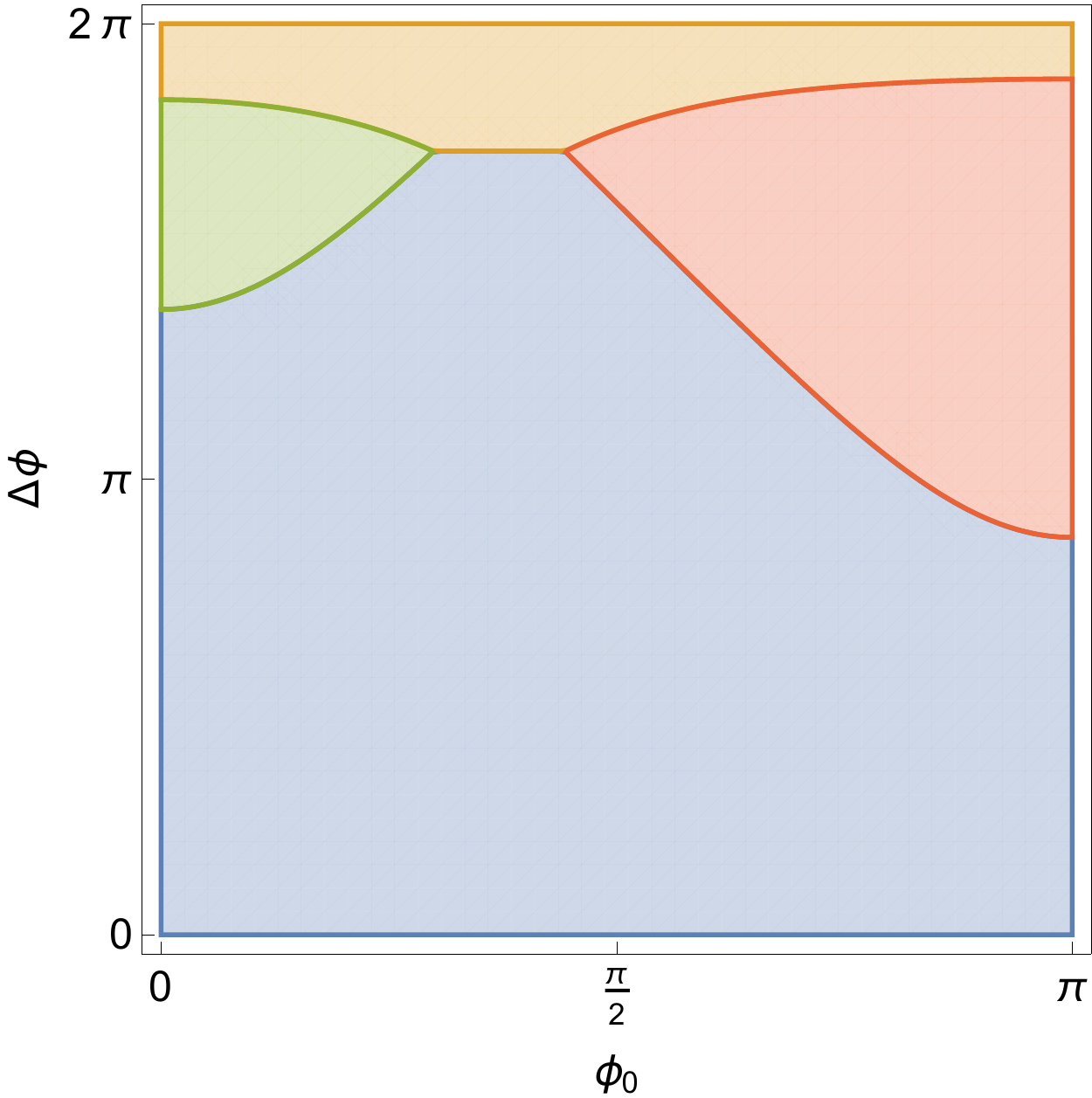}
\caption{$\ell_1=4\pi,\ell_2=2\pi,\ell_3=\pi$}
\end{subfigure}
\caption{Phase diagrams of entanglement entropy for a single interval in region one, with the position of the centre of the interval plotted horizontally and its size vertically. The blue, yellow, green and red regions (also labelled in subfigure (c)) indicate where phases 1, 2, 3 and 4 respectively dominate, as defined in \cref{fig:3wormholePhases}.}
\label{fig:3wormholePhaseDiagram}
\end{figure}

For some specific values of the moduli, phase diagrams of varying interval size and length are shown in \cref{fig:3wormholePhaseDiagram}.
An interesting aspect of these is what they imply for the spatial distribution of entanglement. The Araki-Lieb inequality for the interval and one of the other boundaries is most likely to be saturated when they are closest to one another. Regarding the horizon lengths as proxies for the size of each system, when two boundaries are small and one is large, the small boundaries are not only entangled exclusively with the large one, but actually with a spatially localised region within it. This na\"ive idea of geometric closeness corresponding to entanglement may be a natural guess, but it is striking to see it so precisely realised.

The investigations here should not be viewed as exhaustive, but rather a demonstration of the results that can be quickly obtained from this method. There is much more that can be done in these geometries to understand the entanglement, particularly with measures that are not inherently bipartite, generalising the work in \cite{Balasubramanian:2014hda} to subintervals. One obvious quantity to look at is mutual information between intervals on different boundaries. Further generalisations would be to look at time dependence, allow the wormholes to spin, and to add more boundary regions. There are hints that there is some simple universal behaviour when the parameters $\ell$ become large (which is also a region of moduli space where this phase would be expected to dominate), and the techniques here are well suited to be used to understand this analytically. We leave all such investigations for future work.

\subsection{Torus wormhole}\label{sec:Torus}

The final example we discuss is a black hole with a single exterior, containing a torus hidden behind the event horizon. Once again, we will consider here only a nonrotating version; a spinning generalisation was discussed at length in \cite{Aminneborg:1998si}, and what is done here can be straightforwardly extended to that case. This is a pure state on a single CFT, prepared by a path integral over a torus with a single boundary, a Riemann surface with three real moduli. The field theory interpretation of this state is quite mysterious, but in any case we expect it to be some atypical finely tuned excited state at $t=0$, breaking translational symmetry in both space and time directions.

The partition function is that of a genus two surface, with a reflection symmetry fixing a single circle lying between the handles and splitting the surface into two parts. Apart from the bulk phase we discuss here, there is an infinite family of time-reflection symmetric Euclidean bulks, corresponding to filling in some combination of cycles on one handle of the surface, and the same (reflected) combination of cycles on the other. These are in correspondence with pairs of coprime integers defining the choice of cycle, much like the $SL(2,\ZZ)$ family of 
Euclidean black holes generalising BTZ and Euclidean thermal $AdS$. For these bulks, the Lorentzian spacetime is always just pure $AdS$. Once again we will not worry about which saddle dominates the path integral, but work in the interesting phase with the caveat that the results apply only in certain regions of moduli space.

\begin{figure}
\centering
\includegraphics[width=0.4\textwidth]{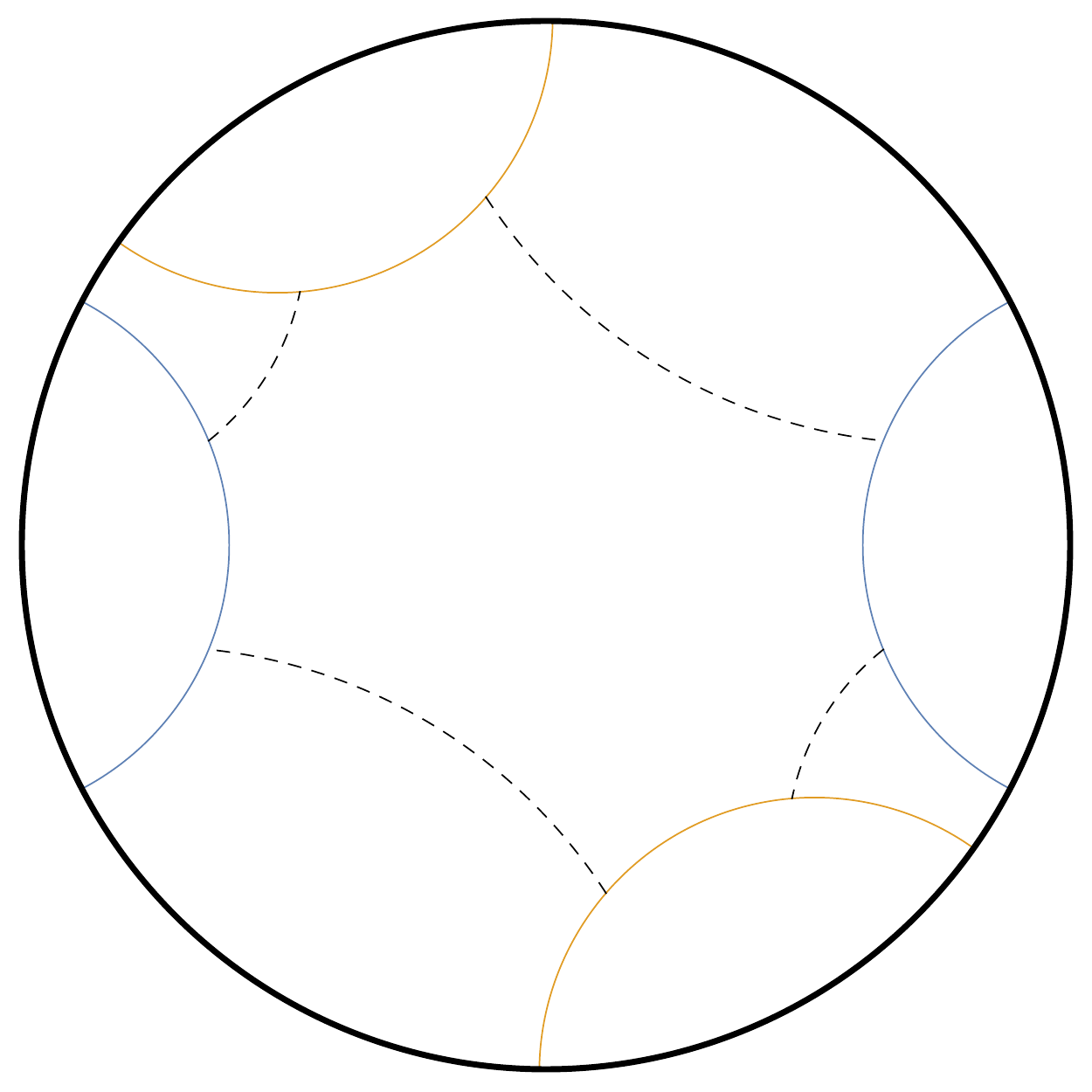}
\caption{The $t=0$ slice of $AdS_3$, showing a fundamental region and event horizon for the torus wormhole. The blue curves are identified by $g$, and the orange by $h$. The event horizon here is the set of dashed geodesics joining the fixed points of $g h g^{-1} h^{-1}$, $h g^{-1} h^{-1} g$, $g^{-1} h^{-1} g h$ and $ h^{-1} g h g^{-1}$.}
\label{fig:torusPoincare}
\end{figure}

As discussed at the beginning of the last section, the geometry comes from a quotient by a free subgroup of $PSL(2,\RR)$ generated by two elements $g$ and $h$. Considering these elements as exponentials of Lie algebra elements in a three dimensional Lorentzian space, they come from a pair of spacelike vectors which may now be boosted to simultaneously have vanishing timelike component. The moduli therefore are the two lengths of the vectors $\lambda>0$ and $\mu>0$, and the angle $\alpha\in (0,\pi/2]$ between them. Picking a convenient basis, $\Gamma$ is generated by
\begin{align}
g&=\exp
\begin{pmatrix}
\lambda & 0 \\
 0 & -\lambda
\end{pmatrix}
=\begin{pmatrix}
e^\lambda & 0 \\
 0 & e^{-\lambda}
\end{pmatrix},\\
h&=
\exp
\begin{pmatrix}
\mu\cos\alpha & \mu\sin\alpha \\
 \mu\sin\alpha & -\mu\cos\alpha
\end{pmatrix}
=\begin{pmatrix}
\cosh\mu+\cos\alpha\sinh\mu & \sin\alpha \sinh\mu \\
 \sin\alpha \sinh\mu & \cosh\mu-\cos\alpha\sinh\mu
\end{pmatrix}.
\end{align}
The resulting geometry is most easily understood once again by looking at the a fundamental region on the $t=0$ Poincar\'e disc, as in \cref{fig:torusPoincare}. This looks very similar to the three boundary case above, except that the identifications are not between neighbouring semicircles but between opposite sides.
To see the topology, this is exactly like the identification of opposite edges of a square to form a torus, except with the corners cut off; this makes a torus with a single asymptotic boundary.

One computational difficulty that we encounter here is that the elements of $\Gamma$ enacting the translation along the boundary by $2\pi$ are not so simple as before. Such `horizon words' are here the conjugates of $g h g^{-1} h^{-1}$; any point on the boundary of the Poincar\'e disc will lie between the fixed points of one such element, which is the exponential of the generator of boundary translations there. This conjugacy class also defines the homotopy class of the event horizon, which is why we call the representatives horizon words. This is exactly like the third asymptotic region in the above three boundary wormhole. Note that the horizon word here is trivial when abelianised, so has trivial homology, since it is the boundary of everything behind the horizon. This is a manifestation of the purity of the state.

Calculating any such word and using equation~\eqref{eq:closedLength} to find its length, we get the horizon radius of the black hole
\begin{equation}
\rp = \frac{1}{\pi} \cosh(2\sin^2\alpha\sinh^2\lambda\sinh^2\mu-1)
\end{equation}
from which it is apparent that we must take the angle between the generators large enough so that $\sin\alpha\sinh\lambda\sinh\mu>1$. As with the three boundary wormhole, this can be understood from the requirement that the circles marking the edges of the fundamental region in \cref{fig:torusPoincare} do not overlap, and ensures that all elements of the group $\Gamma$ are hyperbolic.

To find a nice regularisation to represent boundary points, as before it is simplest to first identify the local Killing vector $\xi$ implementing translations. Since the horizon word is not in a simple form, this requires first inverting a matrix exponential, for example $e^{2\pi\xi}= -g h g^{-1} h^{-1}$, but for hyperbolic $SL(2,\RR)$ matrices, this is a straightforward procedure. We then must choose a reference point, which must be in the right part of the boundary of the Poincar\'e disc, lying on the proper side of the fixed points of $\xi$ so that it is the correct local generator of translations. Since the fixed points move as the parameters vary, there is no universal choice to be made here, but it depends on the parameters. One practical possibility comes from examining the fixed points of $\xi$, being the eigenvectors, and taking the average of the two angles of the resulting vectors; this gives an especially simple answer in the symmetric case $\alpha=\frac{\pi}{2}$. With this initial point chosen, the others may be found by translating with the Killing vector, after which an overall scale can be fixed if desired, by matching to the BTZ result. Note that this will not give boundary representatives in the boundary fundamental region in \cref{fig:torusPoincare}, but a connected fundamental region in the boundary covering space.

Apart from these small extra practical difficulties, the calculations proceed much as before, and we will not reproduce details here. We again examine the entanglement entropy for a single interval on the boundary. It turns out to be much harder in this case to systematically rule out the majority of geodesics built from long words in the generators as always being longer, as was possible for the three boundary wormhole, making a complete characterisation difficult. This is perhaps because the torus behind the horizon may get very `twisted' relative to the generators $g$ and $h$ for some parameters, so that the shortest geodesics are more complicated words in $\Gamma$. Using a different choice of generators for $\Gamma$ may then look more natural, and allow the geometry to be described by some different set of parameters in which the short geodesics are simple words. This is loosely analogous to describing a flat torus with upper half plane modular parameter $\tau$, when short geodesics may look complicated if $\tau$ is not picked in a fundamental region, but instead related by some modular transformation.

In any case, to simplify things we now focus on a special case of the most symmetric geometry, by picking $\alpha=\pi/2$ and $\lambda=\mu$. The one remaining modulus can be tuned to give any value for the horizon radius. The geometry then goes from having no symmetry to a dihedral group, the symmetry group of the square, acting on the boundary as rotations by $\pi/2$ and four possible reflections of the circle.

In this special case, a systematic study of the geodesics is more useful, showing that the more complicated geodesics do not dominate the entanglement entropy of a single interval. We followed a similar method to the three boundary case, of looking at words built from some maximal number of generators over a large sample of different parameters, including for closed geodesics.  Restricting the interval to have its centre lying in a range of angles from $0$ to $\pi/2$, as allowed by the discrete symmetries, and to have length up to $\pi$, as may be done since the state is pure, in fact gives just one nontrivial contributing geodesic, associated to the group element $g$. The three related by the discrete symmetries can be associated with $h$, $h^{-1}g^{-1}h$ and $g h^{-1} g^{-1}$. The entanglement entropy is computed including the closed geodesics in the conjugacy classes $g^{-1},h^{-1},g$ or $h$ respectively (all of equal length here) to satisfy the homology constraint.

\begin{figure}
\centering
\includegraphics[width=0.6\textwidth]{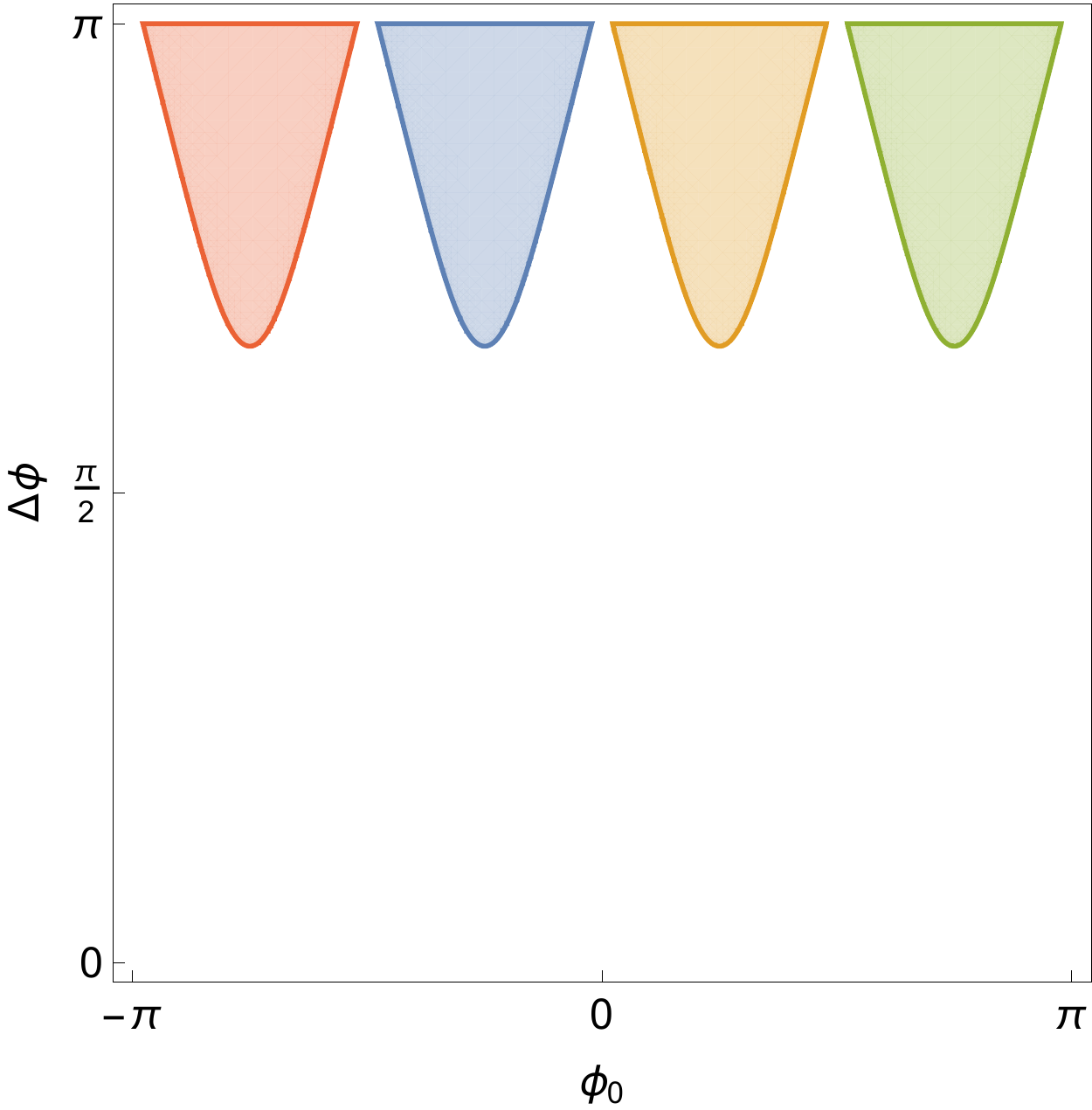}
\caption{Phases of entanglement entropy for a single interval in the symmetric torus wormhole, with horizon radius $\rp=5$. The position of the centre of the interval is plotted horizontally and its size, up to $\pi$, plotted vertically. The four coloured regions correspond to dominance of four different geodesics passing behind the event horizon, all related by the discrete symmetries of the spacetime. The uncoloured region is where the trivial geodesic dominates, giving the BTZ result.}\label{fig:torusWormholePhases}
\end{figure}

These phases dominate when the horizon radius and entangling interval are sufficiently large, and further depend on the position of the interval, as shown in \cref{fig:torusWormholePhases}.
It would be interesting to see if relaxing the symmetry allows for more than these simple phases to dominate, or whether it simply distorts the shape of the regions of dominance.

We do not have any good field theory interpretation for this result. It would be interesting to study in more detail, relax the symmetry assumptions, and to understand what it implies for the state of the field theory. These preliminary results at least show that the answer is not trivial, so entanglement entropy is a useful observable.

\section{Relations between Euclidean and covariant entanglement entropy}
\label{sec:RTHRT}

\subsection{Analytic continuation}
In a Euclidean context, there is a derivation of the Ryu-Takayanagi formula from a Euclidean quantum gravity argument \cite{Lewkowycz:2013nqa}, by performing the replica trick in the bulk Euclidean spacetime. In time-reflection symmetric spacetimes, one may freely pass between Euclidean and Lorentzian descriptions (see \cref{sec:EuclidLorentz}), by analytic continuation, or by a Hartle-Hawking procedure regarding the $t=0$ slice as an initial Cauchy surface, which has zero extrinsic curvature due to the reflection symmetry, allowing for a well-defined Lorentzian evolution. From this, the argument can be regarded as proving the Ryu-Takayanagi prescription for regions lying at such a surface of time-reflection symmetry where the Euclidean description is available, of which static geometries are a special case.

This leaves at least two unanswered questions: firstly, how does the derivation generalise to states with no time-reflection symmetry? It is unclear how to pass to a Euclidean description in such a situation, if possible at all, and we will make no attempt to progress in this direction. The second is whether we can do something weaker, and generalise to time-reflection symmetric geometries, but where the region whose entanglement entropy we would like to compute does not lie at the point of time symmetry.

The natural thing to do here would be to begin with endpoints of an interval on the boundary at $t=0$, which lives naturally in either Lorentzian or Euclidean spacetimes, and to evolve by some Euclidean time $\tau$ in the Euclidean boundary. The Euclidean quantum gravity argument may then be used to justify computing an entanglement entropy for the new interval from a geodesic length in the Euclidean bulk. Finally, to get the result for the real time evolution, assume analyticity and analytically continue the answer to $t=-i \tau$.

It is, however, not in general obvious how to carry out this procedure since the notion of Euclidean time evolution does not appear to be unambiguously defined. This works in BTZ, since there is a unique timelike Killing vector orthogonal to the translation symmetry, and many approaches requiring analytic continuation, such as \cite{Hartman:2013qma}, have relied on this (or equivalently, continuation of a conserved energy). But more complicated Riemann surfaces do not have any continuous symmetries, so this approach does not generalise. Further, it is not clear that the result thus obtained is in fact the length of a geodesic in the Lorentzian spacetime. The technology at hand solves both problems in the present context.

We begin by describing the relevant Euclidean time evolution in the context of a quotient by a Fuchsian group $\Gamma\subseteq PSL(2,\RR)$. Given a point on the boundary of the $AdS$ covering space at $t=0$, described by a singular matrix $p(0)$, there is a translation Killing field of the quotient $\xi\in \mathfrak{sl}(2,\RR)$ defined locally, being in the static case the horizon generator described in \cref{sec:Torus}. Any point on the boundary of the Poincar\'e disc lies between the fixed points of some unique $\gamma\in\Gamma$, and $\gamma=e^{2\pi\xi}$ defines $\xi$. Translation (in either signature) is implemented by $p\mapsto e^{a\xi}\, p\, e^{a\xi^t}$; in Lorentzian signature the orthogonal translation implements time evolutions via $p(t) = e^{t\xi}\, p(0)\, e^{-t\xi^t}$. Motivated by this, continuing to Euclidean signature by $t=-i\tau$, the correct time translation must therefore be $p(\tau) = e^{-i\tau\xi}\, p(0)\, e^{i\tau\xi^t} = e^{-i\tau\xi}\, p(0)\, (e^{-i\tau\xi})^\dagger$; this is a sensible action on the Riemann sphere since the matrix on the right is the conjugate of the matrix on the left. It corresponds to following a geodesic on the Euclidean boundary with the flat metric, orthogonal to the $t=0$ slice, where the metric matches with the flat Lorentzian metric. This is locally uniquely defined, but of course cannot be extended to the whole boundary (except for the torus) consistently with the quotient. An alternative way of characterising it is as the elliptic Killing field on the Riemann sphere with the same fixed points as the generator of translations along the boundary, with an appropriately normalised length.

Now with this notion of Euclidean time evolution, the algebraic expressions for the lengths of the geodesics \eqref{eq:openLength} and \eqref{eq:openLengthEuclid} in the two signatures are manifestly analytic in $t$ or $\tau$, and follow from one another by continuation $t=-i\tau$. With the assumption that entanglement entropies should be appropriately analytic (saving for phase transitions between different geodesics), this could be interpreted as a proof that the covariant proposal \cite{Hubeny:2007xt} follows from the Euclidean calculation in this limited set of circumstances.

This works most straightforwardly in the case that the quotient is by a Fuchsian group $\Gamma$, with the initial points $p(0)$ at $t=0$ lying on the real axis of the Riemann sphere. In this case the time-reflection invariant slice in the bulk is exactly the quotient of the upper half plane, with hyperbolic metric, by $\Gamma$, so the initial geodesic lies on the static slice, as do all the closed geodesics, so the analytically continued lengths all come from real geodesics. However, the Euclidean bulk geometries only follow from Fuchsian groups in a specific phase; most phases come from non-Fuchsian Schottky groups, in which case the Lorentzian bulk is disconnected. For each connected component of the bulk at $t=0$ the Schottky group can by conjugation be put in a form such that the boundaries lie on the real axis. If the initial points lie in the same connected part, and further the initial geodesic can be lifted to covering space in such a way that its endpoints both lie on the real axis, the above argument still goes through, and the Lorentzian geodesic lengths still follow from the appropriately analytically continued Euclidean argument. This captures the fact that by HRT, when the spacetime is disconnected the entanglement entropy (at leading order in $G_N$ at least) becomes disconnected also. Geodesics may only connect points in the same connected component, and the allowed homotopy classes of geodesics are those of the component of spacetime in which they live.

\subsection{Euclidean geodesics without Lorentzian analogues}

From a Euclidean point of view, these restrictions on geodesic endpoints and their allowed homotopy classes seem very much less natural. Even with initial points at $t=0$ and without performing the analytic continuation, viewing the Lewkowycz-Maldacena computation from a purely Euclidean point of view, there is no reason to require that the geodesics (open or closed) lie on the $t=0$ slice. The argument requires the time-reflection symmetry only in order to translate to Euclidean signature; the bulk duals computing R\'enyi entropies may be $\ZZ_n$ replica symmetric without being time-reflection symmetric, with the reflection symmetry restored in the $n\to 1$ limit, leading to surfaces lying away from the time-symmetric slice. From the standpoint of the Lorentzian section, such saddle points would have no geometric interpretation, but would appear as complexified geodesics. This could, for example, allow for nonzero mutual information between intervals in a phase where they lie in disconnected components of the spacetime\footnote{We take the na\"ive point of view of assuming that the least-action saddle always dominates the path integral, but the possibility that it is not on the path of steepest descent should be borne in mind.}.

Complex entangling surfaces have been considered before \cite{Fischetti:2014zja}, but these are not quite the same as what we consider here, which are real geodesics, hence with real length, but living in the Euclidean section of the spacetime. For an explicit illustration, we finish the section with a computation in the low temperature phase of the thermofield double state, to check whether the possibility is realised.

The protocol of \cref{sec:H3} can be employed to quickly compute the lengths of geodesics in Euclidean thermal $AdS$. The spacetime can be described as a quotient of $H^3$ by the group generated by the $PSL(2,\RR)$ transformation with associated M\"obius map $z\mapsto e^{-\beta}z$. This is Fuchsian, but not with respect to a basis with the time reflection implemented by complex conjugation (time reflection here is instead $z\mapsto -1/\bar{z}$), so passing directly to Euclidean signature by putting it in the diagonal subgroup of $AdS$ isometries does not achieve the correct thing here. The boundary circles can instead be chosen at $|z|=1$ and $|z|=e^{\beta/2}$, represented by vectors in $\CC^2$ by
\begin{equation}
\vec{u}_L = \frac{1}{\sqrt{2}}\begin{pmatrix}
e^{i \phi/2} \\ e^{- i \phi/2}
\end{pmatrix}, \quad
\vec{u}_R = \frac{1}{\sqrt{2}}\begin{pmatrix}
e^{\beta/4+ i \phi/2} \\ e^{-\beta/4 - i \phi/2}
\end{pmatrix}.
\end{equation}
From this it is immediate to apply \eqref{eq:openLengthEuclid} to get the lengths of geodesics between points on the same boundary, given by the vacuum $\log \sin^2\left(\frac{\Delta\phi}{2}\right)$ answer, and on opposite boundaries, joined through the torus to get
\begin{equation}
l=\log\left| \sin\left(\frac{\Delta\phi+i\beta/2}{2}\right)\right|^2
= \log\left[ \cosh^2\left(\frac{\beta}{4}\right)\sin^2\left(\frac{\Delta\phi}{2}\right) +\sinh^2\left(\frac{\beta}{4}\right)\cos^2\left(\frac{\Delta\phi}{2}\right) \right].
\end{equation}
This later class of geodesics pass half way round the thermal circle to join boundary points through the Euclidean section, despite them being in disjoint components of the Lorentzian spacetime.
They could give a dominant contribution to the mutual information if they are ever sufficiently short. This has the best chance of happening for large intervals of length $\pi$, at the same angular position on both sides, so there are geodesics passing from $\phi=0$ on one side to $\phi=0$ on the other, and the same between points at $\phi=\pi$. The usual geodesics for the entanglement entropy then have regularised lengths $\log \sin^2\left(\frac{\pi}{2}\right)=0$. The inherently complex geodesic lengths then come from the last equation at $\Delta\phi=0$ to get $\log\sinh^2\left(\frac{\beta}{4}\right)$. The upshot is that these geodesics are short enough to dominate for $\beta < 4\sinh^{-1}1 \approx 3.5$. But if this is the case, we clearly have $\beta<2\pi$, which implies that we are in fact in the high temperature phase where the dominant geometry is the black hole.

In this simple example, it is apparent that the geodesics in the Euclidean space with no Lorentzian counterpart are not the most relevant for computing entanglement entropies. However, it is interesting that this required the extra information about the dominant saddle point geometry, depending on the moduli of the boundary Riemann surface. This can be to some extent intuited from a very na\"ive picture of determining the correct phase, in which the $g$ shortest cycles (for a genus $g$ surface) are filled in. If moving along a geodesic round a nontrivial cycle in the Euclidean direction gives a short answer, then this may indicate that this cycle on the boundary is itself short (being the Euclidean time circle here), and should have been filled in to give a more dominant bulk action.

If the Ryu-Takayanagi prescription is to be regarded as following from the Euclidean quantum gravity replica computation, it is clearly difficult to rule out the possibility of complex geodesics for the Lorentzian prescription, in the precise sense used here. The techniques described here are useful for investigating this further in more involved examples. It would be interesting to know if this phenomenon can be ruled out entirely, or whether it has some physical relevance. If they could be important, and are not just an artefact of the special symmetric situation, then it is crucial to understand what `complex geodesic' means more generally, in spacetimes with no time translation or reflection symmetry.

\section{Discussion}

The main result obtained is a completely algebraic description for regularised lengths of geodesics in solutions to pure gravity with negative cosmological constant, coming from a description of such solutions as quotients. This gives the ability to do such calculations without finding coordinate patches or their overlaps, and without solving a single differential equation, in situations where there are no available symmetries.  The class of geometries this covers includes essentially all pure 3D gravity spacetimes, including states with nonzero angular momentum at the boundary, and nonorientable geometries.

We demonstrated the practical application of the formula for computing entanglement in various states of interest, in many of which an approach using more na\"ive techniques would have been intractable. The first novel results obtained were in the single-exterior $\RR\PP^2$ geon. With access to a single interval covering less than half the boundary circle, there is no difference from the thermal state, though the entanglement entropy of an interval larger than half differs to due to the purity of the state. However, the mutual information between two intervals shows that the dual CFT state is far from being a typical microstate at $t=0$, with highly tuned nonlocal correlations, any given region appearing to be most strongly entangled with regions furthest away from itself. Time evolution quickly destroys these correlations, and it would be interesting to understand better the dynamics of entanglement in this thermalisation.

The next, more involved, example was an entangled state of three noninteracting CFTs, described geometrically by three exterior BTZ regions all joined through a wormhole. Here we find that the entanglement entropy of a single interval is sensitive to the details of the geometry hidden behind the horizon, in particular showing dependence on space and time despite the exterior geometries being symmetric under these translations. There are circumstances where the answer saturates various Araki-Lieb inequalities, which show that entanglement is distributed between the three CFTs in a spatially ordered manner.

The spacetime in the final example is a single exterior black hole with a torus hidden behind the event horizon. Such states with nontrivial hidden topology are particularly mysterious from a field theory standpoint, and our preliminary investigations show that entanglement entropy is a practical and useful observable to analyse them. The results show that even in the most symmetric case, the entanglement entropy of a single interval of sufficient size knows about more than the thermal behaviour, in particular showing the breaking of translation symmetries in space and time.

The surface has only been scratched in the examples here, and there is much more to understand about the physics of entanglement in these geometries.
The ability to compute entanglement entropy efficiently in states with no symmetry, with nonzero momentum, including dynamics, and over a range of moduli provides us more generally with a versatile `laboratory' for expanding our knowledge of the area. We leave more detailed studies to future work.

The final chapter made inroads into obtaining an analytic continuation procedure to obtain HRT from a Euclidean quantum gravity derivation. It would be interesting to understand whether this can be generalised to a less prescriptive set of circumstances, allowing geometries with fields turned on and higher dimensions, as well as relaxing the time-reflection symmetry condition required to translate to Euclidean signature.

Finally, we discussed a very specific way in which HRT might fail, if such a Euclidean quantum gravity derivation is believed. This comes from a spontaneous breaking of the time reflection symmetry when computing the R\'enyi entropies, in such a way that the symmetry is restored in the $n\to 1$ limit. The consequence would be a geodesic in the Euclidean bulk section computing entanglement entropy, with no geometric analogue in the Lorentzian spacetime. It would, further, invalidate the geometric proofs of properties required for consistency, such as strong subadditivity \cite{Wall:2012uf} and causality \cite{Headrick:2014cta}. We describe a simple example where such things turn out to be subdominant, for which the Euclidean version of the geodesic length calculation is well-suited, and it would be interesting to check dominance in more involved cases.

This is related to two other possible problems. One is the assumption of no spontaneous breaking of the cyclic $\ZZ_n$ part of the replica symmetry required in \cite{Lewkowycz:2013nqa}. This is required to make any progress with the argument but it is difficult to rule out and hard to interpret. Perhaps a more basic problem comes from a similar symmetry assumption implicit in the interpretation of the bulk geometries themselves. The Hartle-Hawking procedure requires the time-reflection symmetric slice from which to evolve the Lorentzian geometry, but it may be that the dominant Euclidean saddle point spontaneously breaks this symmetry, so obtaining the dominant Lorentzian geometry becomes impossible. It is known that this never occurs for the torus, but nothing is known at higher genus. A proof that such reflection symmetries are never spontaneously broken would solve all the above problems at a stroke (the dihedral replica symmetry group is generated by the reflections); a counterexample on the other hand would cast much into doubt.

On a more speculative note, the simple and universal form of the algebraic result obtained for entanglement entropy is very suggestive that there may be an alternative way to obtain it directly from conformal field theory. A natural way to attack this problem is in the spirit of previous work \cite{Hartman:2013mia} in vacuum, where the R\'enyi entropies are obtained by correlation functions of twist operators on the sphere, matched to the gravity calculation \cite{Faulkner:2013yia}. The key ingredient is that at large central charge with few low lying operators, there is a universal answer coming from the vacuum block in a conformal block expansion of OPEs, with different geodesics coming from different OPE channels. The first challenge is therefore to classify the OPE channels of operators on higher genus Riemann surfaces, and to obtain a matching with the classification of both the choice of bulk saddle, as well as geodesics in the bulk, including the closed geodesics. This would in itself be very interesting, as it should, for example, give a microscopic interpretation of the homology constraint.

\hrulefill
\paragraph{Acknowledgements}
I would like to thank Mukund Rangamani, Matt Headrick, Simon Ross, Veronika Hubeny and Don Marolf for useful discussions and comments, and the hospitality of the Institute of Advanced Studies, Princeton, and Massachusetts Institute of Technology during the early stages of the project. I am supported by a studentship from STFC.

\bibliographystyle{JHEP}
\bibliography{geons}

\providecommand{\href}[2]{#2}\begingroup\raggedright\begin{thebibliography}{10}

\bibitem{Ryu:2006ef}
S.~Ryu and T.~Takayanagi, {\it Aspects of holographic entanglement entropy},
  {\em JHEP} {\bf 08} (2006) 045,
  [\href{http://xxx.lanl.gov/abs/hep-th/0605073}{{\tt hep-th/0605073}}].

\bibitem{Ryu:2006bv}
S.~Ryu and T.~Takayanagi, {\it {Holographic derivation of entanglement entropy
  from AdS/CFT}},  {\em Phys. Rev. Lett.} {\bf 96} (2006) 181602,
  [\href{http://xxx.lanl.gov/abs/hep-th/0603001}{{\tt hep-th/0603001}}].

\bibitem{Hubeny:2007xt}
V.~E. Hubeny, M.~Rangamani, and T.~Takayanagi, {\it {A Covariant holographic
  entanglement entropy proposal}},  {\em JHEP} {\bf 0707} (2007) 062,
  [\href{http://xxx.lanl.gov/abs/0705.0016}{{\tt arXiv:0705.0016}}].

\bibitem{Banados:1992wn}
M.~Banados, C.~Teitelboim, and J.~Zanelli, {\it {The Black hole in
  three-dimensional space-time}},  {\em Phys.Rev.Lett.} {\bf 69} (1992)
  1849--1851, [\href{http://xxx.lanl.gov/abs/hep-th/9204099}{{\tt
  hep-th/9204099}}].

\bibitem{Banados:1992gq}
M.~Banados, M.~Henneaux, C.~Teitelboim, and J.~Zanelli, {\it {Geometry of the
  (2+1) black hole}},  {\em Phys.Rev.} {\bf D48} (1993) 1506--1525,
  [\href{http://xxx.lanl.gov/abs/gr-qc/9302012}{{\tt gr-qc/9302012}}].

\bibitem{Brill:1995jv}
D.~R. Brill, {\it {Multi - black hole geometries in (2+1)-dimensional
  gravity}},  {\em Phys.Rev.} {\bf D53} (1996) 4133--4176,
  [\href{http://xxx.lanl.gov/abs/gr-qc/9511022}{{\tt gr-qc/9511022}}].

\bibitem{Aminneborg:1997pz}
S.~Aminneborg, I.~Bengtsson, D.~Brill, S.~Holst, and P.~Peldan, {\it {Black
  holes and wormholes in (2+1)-dimensions}},  {\em Class.Quant.Grav.} {\bf 15}
  (1998) 627--644, [\href{http://xxx.lanl.gov/abs/gr-qc/9707036}{{\tt
  gr-qc/9707036}}].

\bibitem{Brill:1998pr}
D.~Brill, {\it {Black holes and wormholes in (2+1)-dimensions}},  {\em
  Lect.Notes Phys.} {\bf 537} (2000) 143,
  [\href{http://xxx.lanl.gov/abs/gr-qc/9904083}{{\tt gr-qc/9904083}}].

\bibitem{Aminneborg:1998si}
S.~Aminneborg, I.~Bengtsson, and S.~Holst, {\it {A Spinning anti-de Sitter
  wormhole}},  {\em Class.Quant.Grav.} {\bf 16} (1999) 363--382,
  [\href{http://xxx.lanl.gov/abs/gr-qc/9805028}{{\tt gr-qc/9805028}}].

\bibitem{Yin:2007at}
X.~Yin, {\it {On Non-handlebody Instantons in 3D Gravity}},  {\em JHEP} {\bf
  0809} (2008) 120, [\href{http://xxx.lanl.gov/abs/0711.2803}{{\tt
  arXiv:0711.2803}}].

\bibitem{Skenderis:2009ju}
K.~Skenderis and B.~C. van Rees, {\it {Holography and wormholes in 2+1
  dimensions}},  {\em Commun.Math.Phys.} {\bf 301} (2011) 583--626,
  [\href{http://xxx.lanl.gov/abs/0912.2090}{{\tt arXiv:0912.2090}}].

\bibitem{Krasnov:2000zq}
K.~Krasnov, {\it {Holography and Riemann surfaces}},  {\em
  Adv.Theor.Math.Phys.} {\bf 4} (2000) 929--979,
  [\href{http://xxx.lanl.gov/abs/hep-th/0005106}{{\tt hep-th/0005106}}].

\bibitem{Maldacena:2001kr}
J.~M. Maldacena, {\it {Eternal black holes in anti-de Sitter}},  {\em JHEP}
  {\bf 0304} (2003) 021, [\href{http://xxx.lanl.gov/abs/hep-th/0106112}{{\tt
  hep-th/0106112}}].

\bibitem{Balasubramanian:2014hda}
V.~Balasubramanian, P.~Hayden, A.~Maloney, D.~Marolf, and S.~F. Ross, {\it
  {Multiboundary Wormholes and Holographic Entanglement}},  {\em
  Class.Quant.Grav.} {\bf 31} (2014) 185015,
  [\href{http://xxx.lanl.gov/abs/1406.2663}{{\tt arXiv:1406.2663}}].

\bibitem{Lewkowycz:2013nqa}
A.~Lewkowycz and J.~Maldacena, {\it {Generalized gravitational entropy}},
  \href{http://xxx.lanl.gov/abs/1304.4926}{{\tt arXiv:1304.4926}}.

\bibitem{Hubeny:2012wa}
V.~E. Hubeny and M.~Rangamani, {\it {Causal Holographic Information}},  {\em
  JHEP} {\bf 1206} (2012) 114, [\href{http://xxx.lanl.gov/abs/1204.1698}{{\tt
  arXiv:1204.1698}}].

\bibitem{Headrick:2014cta}
M.~Headrick, V.~E. Hubeny, A.~Lawrence, and M.~Rangamani, {\it {Causality \&
  holographic entanglement entropy}},
  \href{http://xxx.lanl.gov/abs/1408.6300}{{\tt arXiv:1408.6300}}.

\bibitem{Headrick:2007km}
M.~Headrick and T.~Takayanagi, {\it {A Holographic proof of the strong
  subadditivity of entanglement entropy}},  {\em Phys.Rev.} {\bf D76} (2007)
  106013, [\href{http://xxx.lanl.gov/abs/0704.3719}{{\tt arXiv:0704.3719}}].

\bibitem{Wall:2012uf}
A.~C. Wall, {\it {Maximin Surfaces, and the Strong Subadditivity of the
  Covariant Holographic Entanglement Entropy}},
  \href{http://xxx.lanl.gov/abs/1211.3494}{{\tt arXiv:1211.3494}}.

\bibitem{Haehl:2014zoa}
F.~M. Haehl, T.~Hartman, D.~Marolf, H.~Maxfield, and M.~Rangamani, {\it
  {Topological aspects of generalized gravitational entropy}},
  \href{http://xxx.lanl.gov/abs/1412.7561}{{\tt arXiv:1412.7561}}.

\bibitem{Iizuka:2014wfa}
N.~Iizuka and N.~Ogawa, {\it {On the Entanglement of Multiple CFTs via Rotating
  Black Hole Interior}},  \href{http://xxx.lanl.gov/abs/1402.4548}{{\tt
  arXiv:1402.4548}}.

\bibitem{Hartman:2013qma}
T.~Hartman and J.~Maldacena, {\it {Time Evolution of Entanglement Entropy from
  Black Hole Interiors}},  {\em JHEP} {\bf 1305} (2013) 014,
  [\href{http://xxx.lanl.gov/abs/1303.1080}{{\tt arXiv:1303.1080}}].

\bibitem{Morrison:2012iz}
I.~A. Morrison and M.~M. Roberts, {\it {Mutual information between thermo-field
  doubles and disconnected holographic boundaries}},  {\em JHEP} {\bf 1307}
  (2013) 081, [\href{http://xxx.lanl.gov/abs/1211.2887}{{\tt
  arXiv:1211.2887}}].

\bibitem{Louko:1998hc}
J.~Louko and D.~Marolf, {\it {Single exterior black holes and the AdS / CFT
  conjecture}},  {\em Phys.Rev.} {\bf D59} (1999) 066002,
  [\href{http://xxx.lanl.gov/abs/hep-th/9808081}{{\tt hep-th/9808081}}].

\bibitem{Giulini:1989}
D.~Giulini, {\em 3-manifolds in canonical quantum gravity}.
\newblock PhD thesis, University of Cambridge, 1989.

\bibitem{Friedman:1993ty}
J.~L. Friedman, K.~Schleich, and D.~M. Witt, {\it {Topological censorship}},
  {\em Phys.Rev.Lett.} {\bf 71} (1993) 1486--1489,
  [\href{http://xxx.lanl.gov/abs/gr-qc/9305017}{{\tt gr-qc/9305017}}].

\bibitem{mathematica10}
{Wolfram Research, Inc.}, {\em Mathematica}.
\newblock Wolfram Research, Inc., Champaign, Illinois, version 10.0~ed., 2014.

\bibitem{Araki:1970ba}
H.~Araki and E.~Lieb, {\it {Entropy inequalities}},  {\em Commun.Math.Phys.}
  {\bf 18} (1970) 160--170.

\bibitem{Hubeny:2013gta}
V.~E. Hubeny, H.~Maxfield, M.~Rangamani, and E.~Tonni, {\it {Holographic
  entanglement plateaux}},  {\em JHEP} {\bf 1308} (2013) 092,
  [\href{http://xxx.lanl.gov/abs/1306.4004}{{\tt arXiv:1306.4004}}].

\bibitem{Xi2011}
Z.~Xi, X.-M. Lu, X.~Wang, and Y.~Li, {\it {Necessary and sufficient condition
  for saturating the upper bound of quantum discord}},
  \href{http://xxx.lanl.gov/abs/1111.3837}{{\tt arXiv:1111.3837}}.

\bibitem{Fischetti:2014zja}
S.~Fischetti and D.~Marolf, {\it {Complex Entangling Surfaces for AdS and
  Lifshitz Black Holes?}},  {\em Class.Quant.Grav.} {\bf 31} (2014), no.~21
  214005, [\href{http://xxx.lanl.gov/abs/1407.2900}{{\tt arXiv:1407.2900}}].

\bibitem{Hartman:2013mia}
T.~Hartman, {\it {Entanglement Entropy at Large Central Charge}},
  \href{http://xxx.lanl.gov/abs/1303.6955}{{\tt arXiv:1303.6955}}.

\bibitem{Faulkner:2013yia}
T.~Faulkner, {\it {The Entanglement Renyi Entropies of Disjoint Intervals in
  AdS/CFT}},  \href{http://xxx.lanl.gov/abs/1303.7221}{{\tt arXiv:1303.7221}}.

\end{thebibliography}\endgroup

\end{document}